\documentclass[letterpaper,twocolumn,10pt]{article}
\usepackage{usenix-2020-09}
\PassOptionsToPackage{usenames,dvipsnames}{xcolor}
\usepackage{xcolor}
\usepackage{tikz}
\usepackage{amsmath}

\usepackage{filecontents}

\usepackage{booktabs}
\usepackage{caption}
\usepackage{multirow}
\usepackage{amsmath}
\usepackage{wasysym}
\usepackage[flushleft]{threeparttable}

\usepackage{enumitem}

\usepackage{textcomp}

\usepackage{algpseudocode}
\usepackage{algorithm}

\definecolor{ForestGreen}{RGB}{34, 139, 34}
\algnewcommand{\gComment}[1]{\State%
    \textcolor{ForestGreen}{// #1}}
\newcommand*\circled[1]{\tikz[baseline=(char.base)]{
            \node[shape=circle,draw,inner sep=1pt] (char) {#1};}}

\definecolor{nodeColor}{RGB}{191,191,191}
\newcommand*\nodeParam[1]{\tikz[baseline=(char.base)]{\node[shape=circle,fill=nodeColor,inner sep=0pt, minimum size=5mm] (char) {\textcolor{white}{\textbf{#1}}};}}

\definecolor{nodeRedColor}{RGB}{192,0,0}
\newcommand*\nodeRedParam[1]{\tikz[baseline=(char.base)]{\node[shape=circle,fill=nodeRedColor,inner sep=0pt, minimum size=5mm] (char) {\textcolor{white}{\textbf{#1}}};}}

\usepackage{fancyhdr}
\begin{document}

\date{}

\title{\Large \bf ICSPatch: Automated Vulnerability Localization and Non-Intrusive Hotpatching in Industrial Control Systems using Data Dependence Graphs}

\author{
{\rm Prashant Hari Narayan Rajput\textsuperscript{1}, Constantine Doumanidis\textsuperscript{2}, Michail Maniatakos\textsuperscript{2}}\\\\
\textsuperscript{1}NYU Tandon School of Engineering, Brooklyn, NY, USA\\
\textsuperscript{2}New York University Abu Dhabi, Abu Dhabi, UAE
} 


\maketitle

\begin{abstract}
The paradigm shift of enabling extensive intercommunication between the Operational Technology (OT) and Information Technology (IT) devices allows vulnerabilities typical to the IT world to propagate to the OT side. Therefore, the security layer offered in the past by air gapping is removed, making security patching for OT devices a hard requirement. Conventional patching involves a device reboot to load the patched code in the main memory, which does not apply to OT devices controlling critical processes due to downtime, necessitating in-memory vulnerability patching. Furthermore, these control binaries are often compiled by in-house proprietary compilers, further hindering the patching process and placing reliance on OT vendors for rapid vulnerability discovery and patch development. The current state-of-the-art hotpatching approaches only focus on firmware and/or RTOS. Therefore, in this work, we develop ICSPatch, a framework to automate control logic vulnerability localization using Data Dependence Graphs (DDGs). With the help of DDGs, ICSPatch pinpoints the vulnerability in the control application. As an independent second step, ICSPatch can non-intrusively hotpatch vulnerabilities in the control application directly in the main memory of Programmable Logic Controllers while maintaining reliable continuous operation. To evaluate our framework, we test ICSPatch on a synthetic dataset of 24 vulnerable control application binaries from diverse critical infrastructure sectors. Results show that ICSPatch could successfully localize all vulnerabilities and generate patches accordingly. Furthermore, the patch added negligible latency increase in the execution cycle while maintaining correctness and protection against the vulnerability.
\end{abstract}

\section{Introduction} \label{sec:introduction}
Critical infrastructure such as water treatment plants, the electric grid, chemical manufacturing, and many more rely on various control systems/components broadly identified as Industrial Control Systems (ICS) for regulating physical processes based on industrial logic, necessary for reliable and uninterrupted operation of deeply interconnected critical infrastructure~\cite{art:ics_nist}. Traditionally, these ICS devices executed only the industrial logic, remained confined to the industrial network, and often used proprietary software. However, with the advent of Industry 4.0, the industrial internet of things (IIoT) has gained popularity due to its real-time visibility into the industrial process for robust, affordable asset monitoring and rapid diagnostics while requiring minimal human interaction for data exchange~\cite{art:industry_four}. There is also a trend of shifting away from proprietary software and embracing open source since vendors do not need to reinvent the wheel and develop SSH or web server implementations from scratch. However, these advances in industrial computation also transfer vulnerabilities and attack vectors from the Information Technology (IT) sector to the industrial domain.

As a result, attacks on critical infrastructure have become commonplace in the past few years. These attacks often exploit a vulnerability in the IT infrastructure to reach the Operational Technology (OT) control systems, leading to devastating consequences. One such prominent attack is Stuxnet, which altered the code of the industrial logic running on Siemens PLCs~\cite{art:stuxnet}. Some other examples include a self-replicating virus on the computer network of Saudi Aramco, which shut down 5.7 million barrels of output per day - more than 5\% of the global oil supply~\cite{art:saudi_aramco}. Another widely known instance is the 2015 Ukraine blackout, which involved an attack on the power grid infrastructure that left approximately 225,000 customers without electricity for several hours~\cite{art:ukraine_blackout}. Alternatively, a much more sophisticated attack in December 2016, dubbed Industroyer, with support for four different industrial communication protocols and denial-of-service (DoS) attack implementation for Siemens SIPROTEC family of protection relays, led to another power outage~\cite{art:eset_industroyer}. As a result, organizations have increased their budgets to reduce the risk of cyberattacks. In an investigation carried out by SANS on 340 professionals, 42\% of the respondents confirmed an increase in the cybersecurity budget for their organization~\cite{art:ics_budget}.

A typical ICS setup consists of several control loops utilizing sensors, actuators, and a controller to regulate physical processes. The control logic developed for managing the industrial process executes inside a runtime (privileged software to manage proper application execution, handling I/O, debugging, and more) and is compiled from IEC 61131-3 programming languages. Recently, vendors producing PLCs have started embracing embedded Linux with a real-time patch \cite{misc:simenes_yocto, misc:wago_pfc_sdk, misc:janztec_linux, misc:codesys_device_list}. Choosing Linux as the OS in PLCs enables the utilization of open-source libraries such as OpenSSL, Apache HTTP server, and jQuery, to name a few. A prime example of this trend is the Codesys platform, currently utilized by at least 80 industrial device vendors and supports more than 400 devices from a wide variety of manufacturers~\cite{misc:codesys_device_list}. Four top PLC vendors, namely Mitsubishi, Schneider Electric, Bosch Rexroth, and ABB, hold $\approx$28\% of the global PLC market share~\cite{misc:statistaPLCmarket} and integrate Codesys into their products. Furthermore, the Codesys ecosystem is growing, with high-profile vendors adopting it in their newer product lines to reduce maintenance efforts for their increasingly complex ecosystems \cite{misc:adoption-mitsubishi, misc:adoption-br}, and minor vendors are using it as a turn-key solution to lower the barrier of entry into the market.

Most of the research in the domain focuses on obtaining guarantees about the functionality of the PLC and discovering vulnerabilities: Formal verification of the PLC software~\cite{art:formal_verification_1, proc:formal_verification_2, proc:state_1}, security evaluations~\cite{proc:sec_eval_1, proc:sec_eval_2, art:sec_eval_3}, reverse engineering~\cite{art:icsref}, and fuzzing~\cite{proc:icsfuzz}. These vulnerabilities can result from coding errors (incorrect array indexing) or inherent issues in system libraries (missing bound checks), leading to exploitable vulnerabilities in the control application. Such vulnerabilities can impact not only the control binary itself but also the runtime (i.e., its supervisory software), causing undesired process performance or a complete DoS attack requiring a reboot.

Similar to regular system updates, preventing the exploitation of control application binaries due to vulnerabilities in their code requires patching. However, it remains unexplored in the ICS domain, which has unique constraints. For instance, PLCs regulate critical physical processes and do not allow for any downtime or intrusive modifications to the control application that might affect the continued operation of critical infrastructure, making hotpatching essential. Moreover, control application binaries are non-executable binaries compiled by a proprietary compiler and run on a custom runtime, such as the Codesys runtime system presented earlier. Such nuances prevent the applicability of conventional patching solutions and require a more tailored approach to automated vulnerability localization and hotpatching without interrupting the execution of control logic. In literature, patching ICS devices has predominantly focused only on the firmware, as evident in HERA, which utilizes hardware-based debugging features in Cortex-M microcontrollers to patch vulnerabilities in FreeRTOS~\cite{proc:hera}, and RapidPatch, which employs eBPF virtual machines to execute standard patches released by RTOS developers on resource-constrained devices~\cite{proc:rapid_patch}. On the other hand, further related work on this topic overlooks the patching element and focuses instead on ensuring continuity of operation, employing hot standby redundant PLCs for various purposes such as resiliency against cyberattacks, validating sensor data, process continuity, and improving reliability~\cite{art:redundancy_recovery_1, art:redundancy_recovery_2, proc:redundancy_recovery_3, proc:redundancy_recovery_4}. Nevertheless, automated vulnerability hotpatching is a trending topic in other domains like in Android for performing multi-level adaptive patching~\cite{proc:adaptive_live_patching, proc:anroid_patching_2}, and in open-source software~\cite{proc:opensource_patching_1, proc:kernel_patching_1}, but it remains unexplored in the ICS domain.

This paper proposes ICSPatch, a novel approach for automated vulnerability localization and non-intrusive in-memory hotpatching. The methodology starts with capturing an execution snapshot of the development/test PLC device, rehosting it in an \texttt{angr} instance running on a host for discovering and localizing vulnerabilities using data dependency graphs. Then, we utilize this information to calculate relevant offsets, create a patch for the vulnerability, and non-intrusively patch them in the main memory of the deployed PLC executing the control application. To our knowledge, this work is the first attempt at automated vulnerability localization and hotpatching of non-executable~\emph{ICS control binaries}. In summary, our contributions are the following:
\begin{itemize}[nosep, leftmargin=0.05in]
\item We develop a methodology for automated vulnerability localization for non-executable control application binaries using custom-built data dependence graphs. Our method does not need access to the source code or binary instrumentation; it only needs a copy of the original unmodified control binary file.

\item In parallel, we develop a methodology for control binary hotpatching that can be performed with only remote root access to the device and without any hardware support. Our hotpatching is non-intrusive and decoupled from our automated vulnerability localization. It can be used if there exists no other way of patching the vulnerability. 

\item We create a diverse synthetic dataset of 24 vulnerable control application binaries from five critical infrastructure sectors using the top entries of the latest Common Weakness Enumeration list.

\item We evaluate ICSPatch on: 1) Our synthetic dataset, 2) Other state-of-the-art datasets \cite{proc:icsfuzz}, 3) A hardware-in-the-loop desalination plant model employing a real PLC executing a vulnerable control binary.
\end{itemize}

\section{Preliminaries} \label{sec:preliminaries}
We focus on PLCs that employ Linux as the base operating system and runtime for executing control logic. For the current version of ICSPatch, we target the~\emph{Codesys} Runtime, a multi-platform execution environment employed for programming controllers in IEC 61131-3 compliant programming languages. It operates predominantly on PLCs running a lightweight version of RT-patched Linux OS and comes packed in a self-contained ELF binary that spawns threads for various components. These components include \texttt{MainTask}, which handles the execution of the control application binary. The runtime is not a Position Independent Executable (PIE), so its memory regions are mapped to a fixed location. Application binary execution happens in the three stages that comprise the scan cycle: 1) I/O peripheral inputs sampling. 2) Execution of the control logic. 3) Writing outputs to the I/O peripherals. Since \texttt{MainTask} executes the control application as a thread, they share process memory space. Thus, changes from the control application can impact the runtime. Codesys compiler compiles the control logic into a proprietary format that is then mapped into memory by the runtime. The runtime also initializes the internal address table that stores absolute addresses of the various control application, runtime, and shared library functions.

\begin{figure}[t]
    \centering
    \includegraphics[width=\columnwidth, trim={0.6cm 1.4cm 9.7cm 8.6cm}, clip]{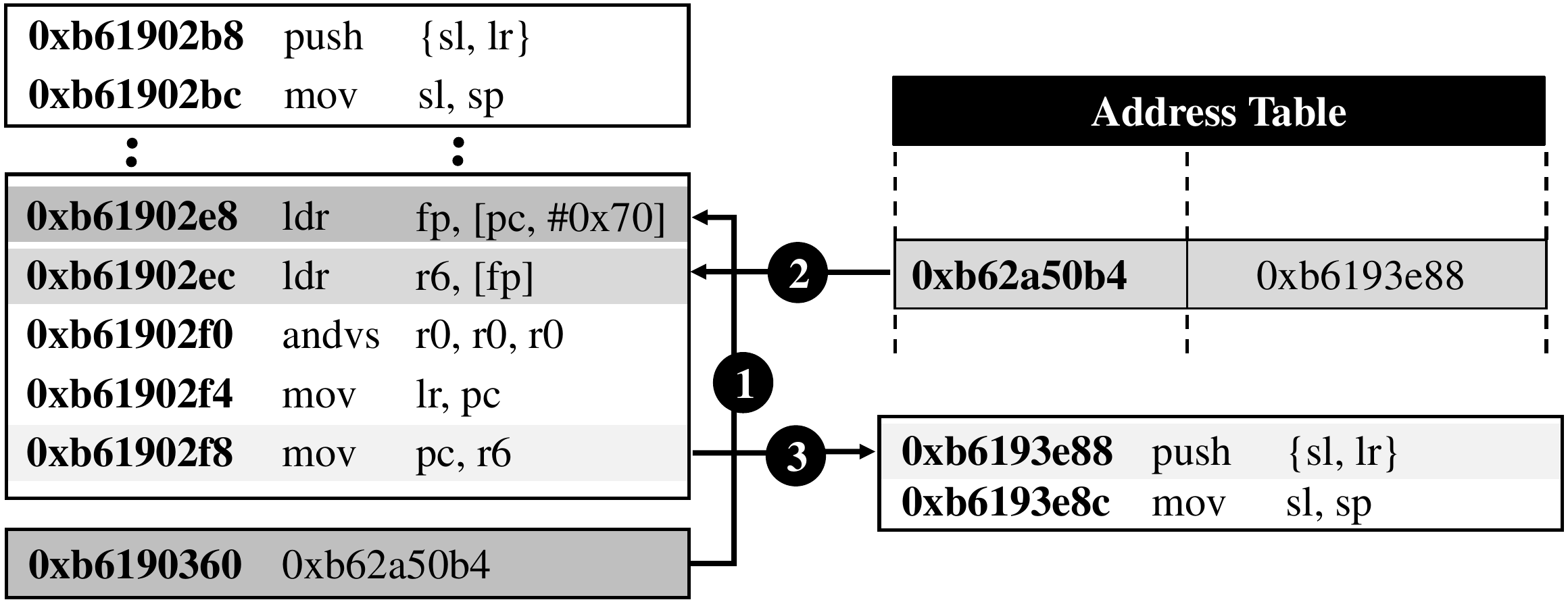}
    \captionof{figure}{Control application directly modifying the \texttt{PC} to branch to the next function in Wago PFC100 PLC.}
    \label{fig:codesys_wago_exec_flow}
    \vspace{-0.4cm}
\end{figure}

Figure~\ref{fig:codesys_wago_exec_flow} shows the steps involved in branching to a called function, 1) The program loads the base address of the address table using PC-relative addressing into \texttt{FP} (\texttt{R11}), 2) it then loads the absolute address of the next function in \texttt{R6}. 3) The \texttt{LR} register is overwritten with the return address, and then the \texttt{PC} is updated with the value of the \texttt{R6} register, currently holding the address of the next function. Traditional approaches exchange parameters between functions using both registers and memory locations. However, function parameters are passed only through memory locations in control application binaries compiled by the Codesys compiler. Codesys runtime also utilizes shared functions provided by standard C libraries (e.g., \texttt{Libc}, \texttt{Libm}), using a wrapper function to move the parameters from memory to appropriate registers.

\begin{figure}[t]
    \centering
    \includegraphics[width=\columnwidth, trim={1.8cm 7cm 13.1cm 0.1cm}, clip]{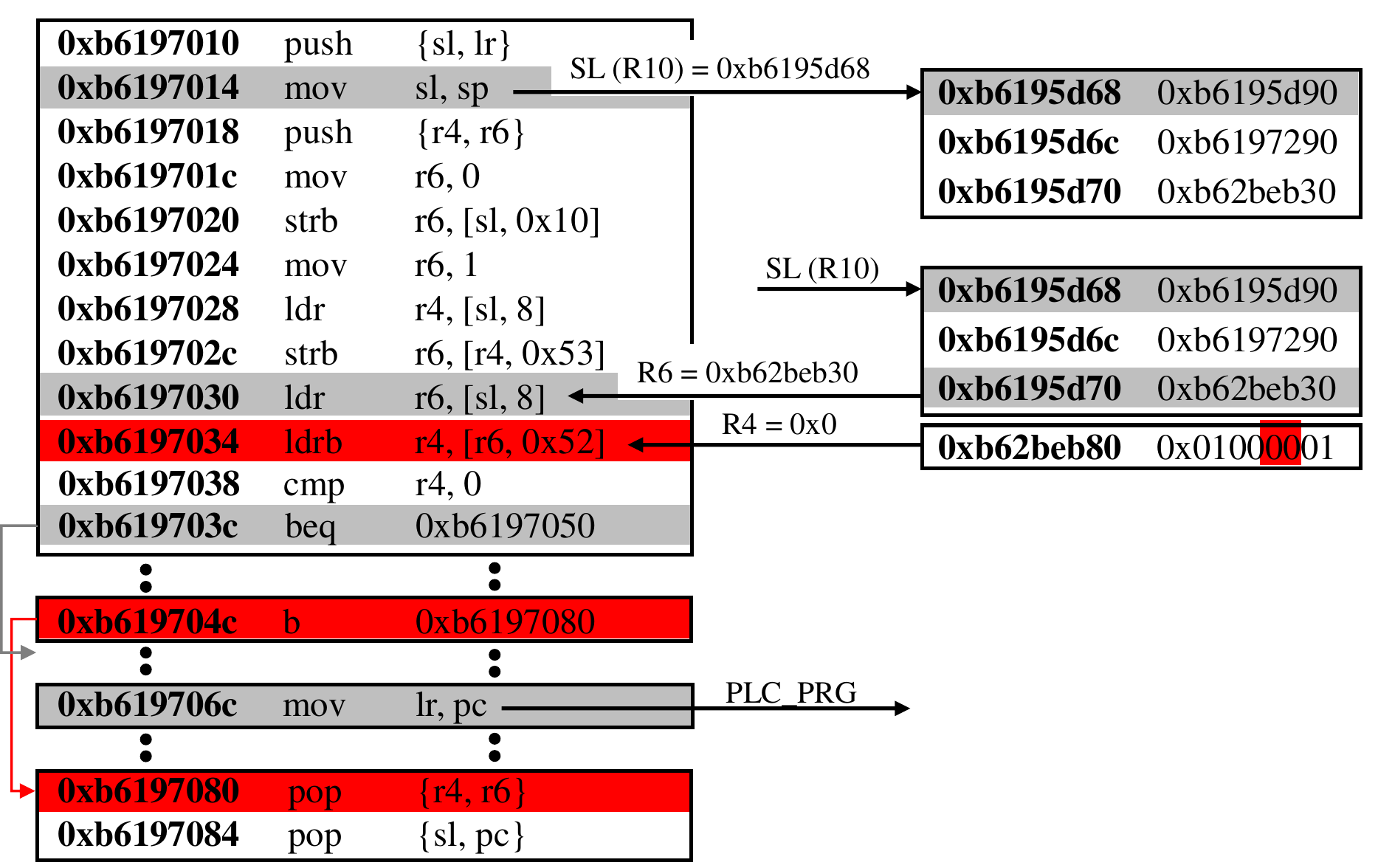}
    \captionof{figure}{Runtime function executed just before the execution of the entry point of the control application.}
    \label{fig:motivating_example}
\end{figure}

Unlike GCC, which employs a Stack Smashing protector, the proprietary Codesys compiler does not implement such protections, enabling malicious writing on the stack. This means that an adversary can control the state of the runtime (the supervisory software for the control application) by exploiting vulnerabilities in the control application itself. To illustrate this, as a motivating example, consider the loader function for the control application, depicted in Figure~\ref{fig:motivating_example}, that passes control to \texttt{PLC\_PRG} (the equivalent of \emph{main} in C) if the byte extracted from the address \texttt{0xb62beb82} is zero. By exploiting vulnerabilities in the control logic, an adversary can manipulate this byte to skip the execution and output phases of the scan cycle, essentially replaying the unprocessed previous written values in the memory-mapped I/O registers.

\section{Problem Formulation} \label{sec:problem_formulation}
In this work, we explore the security implications to the PLC due to vulnerabilities in control applications and the methodology to locate and patch them. More specifically, we endeavor to solve these research problems:

\noindent \textbf{Automating vulnerability localization in the non-executable ICS binary.} ICS devices operate in the field continuously without being taken offline, which only occurs during the scheduled maintenance periods, making hotpatching the only feasible option to patch critical vulnerabilities. This technique patches directly in the main memory during the execution of a program, requiring \emph{precise} knowledge of the location of the vulnerability. The current state-of-the-art patching techniques, such as HERA~\cite{proc:hera} and RapidPatch~\cite{proc:rapid_patch}, assume access to the patched binary or the source code to identify the precise location of the patch, eliminating the localization issue. In our scenario, we assume only access to the binary currently controlling the industrial process, making localization more challenging.

At the same time, our focus is to patch \emph{only} the control application while excluding any of its surrounding logic to avoid unintended system consequences. Vulnerabilities in the control application can manifest from runtime imported functions. For instance, a control application utilizing the \texttt{MemSet} function (to initialize memory for a variable) without implementing proper bound checks will crash the runtime while implementing a store (\texttt{str}) instruction inside the \texttt{MemSet} function. So, the application does not crash the runtime but sends unexpected parameters to the \texttt{MemSet} function that leads to the crash. However, these shared runtime functions may also be used by other parts of the code with benign impact. Therefore, we aim to patch the vulnerability \emph{in the control application} rather than the imported functions; thus, localizing the vulnerability in the control binary is critical.

\noindent \textbf{Software-only remote hotpatching of vulnerable control applications.} In general, hotpatching involves modifying the execution flow in the control logic towards the implemented trampoline function by implementing a hook in the original assembly code. Codesys does not provide hotpatching functionality. The development environment (IDE) allows forcing variable value, which writes the variable value in each cycle, permanently holding the variable at the forced value. However, any changes to the code recompiles and redownloads the updated control application binary to the PLC, which is not desirable during continuous operation. Current state-of-the-art solutions such as HERA~\cite{proc:hera} and RapidPatch~\cite{proc:rapid_patch} patch RTOS compiled like a traditional binary and cannot directly patch non-executable control applications. Furthermore, HERA utilizes Flash Patch and Breakpoint Unit (FPB) on Cortex M3/M4 for dynamically patching embedded devices. Such features limit the applicability of these patching techniques on devices with other microcontrollers and require hardware support. In our scenario, the PLC cannot be decommissioned to add hardware support. Therefore, we aim to apply the patch by remotely connecting to the PLC with admin privileges, a realistic and prevalent way to connect to a PLC in the field.

\subsection{Threat Model and Assumptions}\label{sec:threatmodel}
We consider a field device, a PLC that regulates a physical industrial process, connected to the industrial control network to enable loading, monitoring, and managing the control logic through a Human Machine Interface (HMI). This PLC also receives sensor inputs processed by the currently executing control logic to produce appropriate outputs relayed to the actuators, allowing the PLC to impact the physical state of the industrial process based on the implemented logic. An intelligent adversary can perform man-in-the-middle (MITM) attacks by intercepting and modifying communication with the critical device in the following ways to deliver malicious inputs to the target PLC: 1) Manipulating the sensor inputs to the PLC by compromising the sensor~\cite{proc:limiting_ics} or by changing the values in the memory-mapped I/O registers~\cite{proc:icsfuzz}. 2) Intercepting and modifying the network packets sent from the HMI to the currently executing control logic, enabling data modification in the program while injecting false data to the HMI~\cite{art:icsref}.

We assume that the adversary is limited to data injection/modification attacks by manipulating the sensors or the network traffic. Because if the adversary can modify the control logic, the patch can be removed/overwritten. Such scenarios require orthogonal control logic modification protection solutions, for instance, using checksums, digital signatures~\cite{misc:wibu}, control logic comparison while uploading~\cite{art:scada_pccc}, and formal verification employing behavior~\cite{art:behavior_1, art:behavior_2, proc:sec_eval_1, proc:behavior_4}, state~\cite{proc:state_1, proc:state_2, art:state_3}, specification verification~\cite{proc:spec_1, art:spec_2, art:spec_3}. 

Since ICSPatch targets hotpatching of control applications instead of the firmware, we cannot assume the existence of patches from a trusted source (an assumption by HERA~\cite{proc:hera} and RapidPatch~\cite{proc:rapid_patch}). Instead, ICSPatch creates its patches by populating simple skeleton patches with concrete values. To identify and locate the vulnerability, ICSPatch assumes the existence of at least one exploit input, a typical assumption made in literature~\cite{proc:vuln_localization}. For our target Codesys platform control binaries, we use ICSFuzz~\cite{proc:icsfuzz} to identify a single exploit input. However, similar to existing approaches in literature~\cite{proc:instaguard, proc:spider, proc:adaptive_live_patching}, ICSPatch cannot guarantee the correctness of the generated patch, which should be decided by the process engineers on a development PLC. However, it ascertains the safety of the generated patch similar to RapidPatch~\cite{proc:rapid_patch}, elaborated further in Section~\ref{sec:patchgen}.

\section{ICSPatch Methodology}
\begin{figure*}[t]
    \centering
    \includegraphics[width=\linewidth, trim={0.1cm 6.7cm 0.1cm 2cm}, clip]{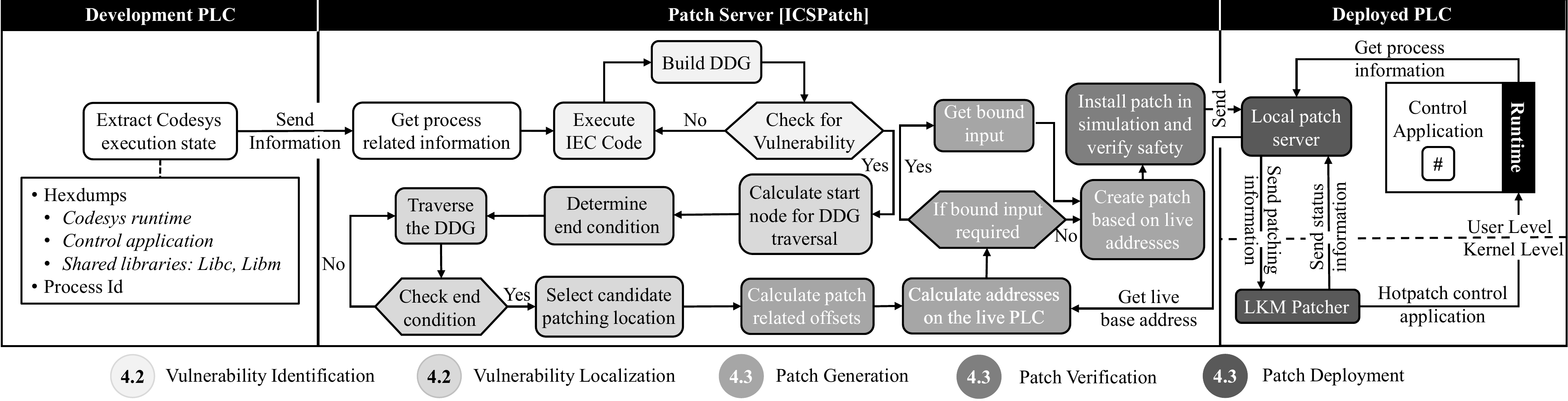}
    \captionof{figure}{ICSPatch system design.}
    \label{fig:icspatch}
    \vspace{-0.4cm}
\end{figure*}

Figure~\ref{fig:icspatch} displays the overall workflow involved with patching control application binaries. The process begins with assuming access to an exploit input that can crash the control application executing on the development PLC, an assumption already made in the literature for automated vulnerability localization~\cite{proc:vuln_localization}. Specialized tools such as ICSFuzz~\cite{proc:icsfuzz} enable the fuzzing of control application binaries, providing the exploit input if available. ICSPatch extracts hexdumps of the runtime process memory space, the \texttt{MainTask} thread executing the control application, and any other required shared libraries. For example, some of the vulnerable control logic in our experiments use shared functions from \texttt{Libc} and \texttt{Libm}. We ensure that the exploit input is stored in the memory-mapped I/O registers before extracting the hexdumps. ICSPatch also extracts other information, such as critical addresses, to enable rehosting of the control application.

\subsection{Overview}
\begin{algorithm}[hbt!]
\caption{DDG-based vulnerability localization and patch generation.}\label{alg:icspatch_full_mod}
\begin{algorithmic}[1]

\scriptsize

\Require $hexdumps$ 
\gComment{Initialize simulation, get simulation state and instruction program counter}
\State $sim, state, pc \gets$ INIT\_SIM($hexdumps$)
\State $end \gets$ CALC\_END\_ADDR($start, state$)
\gComment{Enable vulnerability detection rules in simulation}
\State $sim$.ENABLE\_DETECTION\_RULES()
\State $ddg \gets$ INIT\_DDG()

\While{$pc \neq end$}\label{alg:icspatch_full:while}

    \gComment{Add instruction node for current instruction, containing its operands}
    \State ddg.ADD\_INSTR\_NODE($pc, state.pc.oprnds$)

    \If{$state.op$ = `mem\_write'}
        \gComment{If sim writes to memory, add mem node and connect it to the instr node}
        \State ddg.ADD\_MEM\_NODE($state.mem\_write\_addr$)
        \State ddg.ADD\_EDGE($pc, state.mem\_write\_addr$, `stores')
    \ElsIf{$state.op$ = `mem\_read'}
        \gComment{If sim reads from memory, add mem node and connect it to instr node}
        \State ddg.ADD\_MEM\_NODE($state.mem\_read\_addr$)
        \State ddg.ADD\_EDGE($state.mem\_read\_addr, pc$, `loads')
    \ElsIf{$state.op =$ `reg\_write'}
        \gComment{If write to reg, connect instr node to previous reg state (transition node)}
        \State ddg.ADD\_EDGE(PRV\_REG\_STATE($state.pc.oprnd2$), $pc$, `next')
    \EndIf

    
    \gComment{Detect vulnerability using memory violation rules}
    \If{DETECT\_VULNERABILITY($state$)}
        \gComment{Locate DDG traversal starting point}
        \State $start\_addr \gets$ GET\_COMPARISON\_INSTRUCTION($state.block$)
        \gComment{Get code block bounds for DDG traversal}
        \State $block\_start, block\_end \gets$ GET\_NEAREST\_APP\_BLOCK\_ADR()

        \gComment{Traverse DDG using DFS algorithm to get patch address}
        \State $sim\_p\_addr \gets$ DFS($ddg, start\_addr, block\_start, block\_end$)
    
        \gComment{Check if patch address is valid}
        \If{CHECK\_RANGE($sim\_p\_addr, block\_start, block\_end$)is false}
            \State FAIL()
        \EndIf
        \State $b\_addr \gets$ GET\_BASE\_ADDR()
        \gComment{Create patch based on simulation and deployed PLC information}
        \State$patch,hook,liv\_p\_addr\gets$ BUILD\_PATCH($state, sim\_p\_addr, b\_addr$)
        \gComment{Deploy patch by sending it to the local patch server on the PLC}
        \State DEPLOY\_PATCH($patch, hook, liv\_p\_addr$)
        \State EXIT()
    \EndIf

    \State $state, pc \gets sim$.SIM\_STEP()
\EndWhile

\end{algorithmic}
\end{algorithm}

As shown in Algorithm~\ref{alg:icspatch_full_mod}, having extracted all the relevant information and memory snapshots from the development PLC, ICSPatch starts by initiating an~\texttt{angr} simulation (an open-source binary analysis platform combining static and dynamic analysis techniques~\cite{misc:angr}). The simulation instance is initialized with the memory hexdumps by \texttt{INIT\_SIM()}, which also sets the \texttt{PC} to the start address of the function executed just before the beginning of the PLC control process: ``\texttt{PLC\_PRG}'',  (also shown in Figure~\ref{fig:motivating_example}). After the end address of the simulation is calculated using \texttt{CALC\_END\_ADDR()}, ICSPatch calls \texttt{ENABLE\_DETECTION\_RULES()} to enable memory violation detection rules based on attack observations. This functionality leverages breakpoint support in the \texttt{angr} execution engine. Starting at line \ref{alg:icspatch_full:while}, we step through the simulation performing \emph{concolic} execution (concrete input combined with symbolic execution) of the loaded control application. As ICSPatch executes the control logic in \texttt{angr}, it iteratively builds a custom Data Dependence Graph (DDG) using \texttt{ADD\_INSTR\_NODE()}, \texttt{ADD\_MEM\_NODE()}, and \texttt{ADD\_EDGE()} to add instruction, memory and transition type nodes, respectively (discussed further in Section \ref{sec:vuln_id}). Upon detecting a vulnerability in the current control application, ICSPatch performs localization by traversing the DDG. Traversal happens in \texttt{DFS()} starting from the node calculated by \texttt{GET\_COMPARISON\_INSTRUCTION()} and is bounded by the addresses of the control application function block closest to the vulnerability manifestation, as calculated by \texttt{GET\_NEAREST\_APP\_BLOCK\_AD()}. After completing this step, ICSPatch prompts the user to select the appropriate patching candidate if multiple options are available. In our dataset, this occurs only for \texttt{MemUtils.BitCpy} as it can be exploited by manipulating either the \texttt{size} or \texttt{offset} parameters, having two viable starting nodes for vulnerability localization. However, ICSPatch provides the user with the malicious input at the identified memory location, which can be effortlessly verified with the exploit input to choose the correct candidate. Once the patching candidate (memory location) is selected, ICSPatch calls \texttt{GET\_BASE\_ADDR()} to gather important base address information of the runtime from the deployed PLC. Consequently, it invokes \texttt{BUILD\_PATCH()} combining the offsets calculated from the \texttt{angr} instance with the live base address information to build the patch and calculate all appropriate offsets, such as the offset for the memory location of the patch, the hook, and much more. Finally, in \texttt{DEPLOY\_PATCH()}, ICSPatch communicates the patching information to the local patch server running on the deployed PLC, which relays it to the Loadable Kernel Module (LKM) patcher, patching the control application.

\subsection{Step 1: Vulnerability Identification \& Localization}\label{sec:vuln_id}
\begin{figure*}[t]
    \centering
    \includegraphics[width=\textwidth, trim={0.9cm 15.7cm 4.3cm 1cm}, clip]{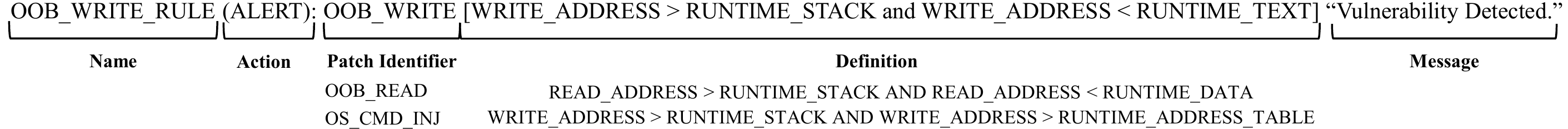}
    \captionof{figure}{Ruleset format for ICSPatch with an example.}
    \label{fig:icspatch_example_ruleset}
    \vspace{-0.4cm}
\end{figure*}

\noindent \textbf{Vulnerability Identification.} ICSPatch performs vulnerability identification in \texttt{DETECT\_VULNERABILITY()} by utilizing the breakpoint functionality in the execution engine of \texttt{angr}. For executing control logic, the runtime passes execution to the control application, whose initial cycle sets up data structures, such as the address table, utilized in consecutive executions. So, in the case of out-of-bounds write and read for our target control applications, the impact of the vulnerability can remain isolated to the control application or even impact the runtime. ICSPatch handles these two cases of impacts differently in \texttt{ENABLE\_DETECTION\_RULES()}. For malicious reads and writes, ICSPatch reviews all the store and load operations while executing the control application in \texttt{angr} to verify if any of them write below the highest stack address of the control application. On the other hand, malicious out-of-bounds write/read can also have a limited impact on the control application. For detecting such conditions, ICSPatch executes the control application with legitimate inputs to capture the offset of stack writes for each function inside their corresponding stack frames, later comparing it with the writes of the control application with exploit input. Since we assume that the adversary only has access to the input of the control application, the changes in the memory locations written by the currently executing control logic are only the result of the exploit input. However, the absolute write/read addresses can change; therefore, ICSPatch captures the offset of a memory location written or read per stack frame for each function.

Finally, OS command injection is a unique case of out-of-bounds write involving writing to the address table with the address of the malicious payload, effectively redirecting execution flow toward the payload. ICSPatch monitors store operations on the address table while executing the control application in \texttt{angr}. The address table is initialized once (in the first scan cycle) and only changes when loading a new control application or restarting the runtime.

It should be noted that vulnerability identification in ICSPatch only recognizes known categories of vulnerabilities in unknown compiled application binaries because it relies on a core set of rules for detecting these vulnerabilities (defined in \texttt{ENABLE\_DETECTION\_RULES()}). So, expanding the ruleset enables the detection of more vulnerabilities.

\textbf{Vulnerability ruleset example.} Memory operations on the control application stack should remain isolated from the runtime stack so that the runtime can continue operating uncorrupted. Furthermore, considering Figure~\ref{fig:stack_iec_vulnerabilities}, the control application stack is followed by the runtime stack, control and runtime code, the address table, and the data section. The control application, during its execution, does not need to manipulate any of these memory regions, so a vulnerability ruleset to capture malicious out-of-bounds write requires checking for memory write operations at an address greater than the runtime stack address as shown in Figure~\ref{fig:icspatch_example_ruleset}. Other components of the ICSPatch ruleset include a~\emph{name}, a~\emph{message} printed upon detecting a particular vulnerability, and the corresponding~\emph{action} (\texttt{WARN} or~\texttt{ALERT}). The~\emph{patch identifier} specifies the corresponding patch for utilization. Finally, the~\emph{rule definition} specifies a logical relationship for vulnerability detection. It can utilize hardcoded values or special addresses dynamically identified by ICSPatch, such as~\texttt{WRITE/READ\_ADDRESS},~\texttt{RUNTIME\_STACK/DATA}, and~\texttt{RUNTIME\_ADDRESS\_TABLE}.

\begin{figure}[t]
    \centering
    \includegraphics[width=\columnwidth, trim={2cm 6.2cm 6.5cm 1.7cm}, clip]{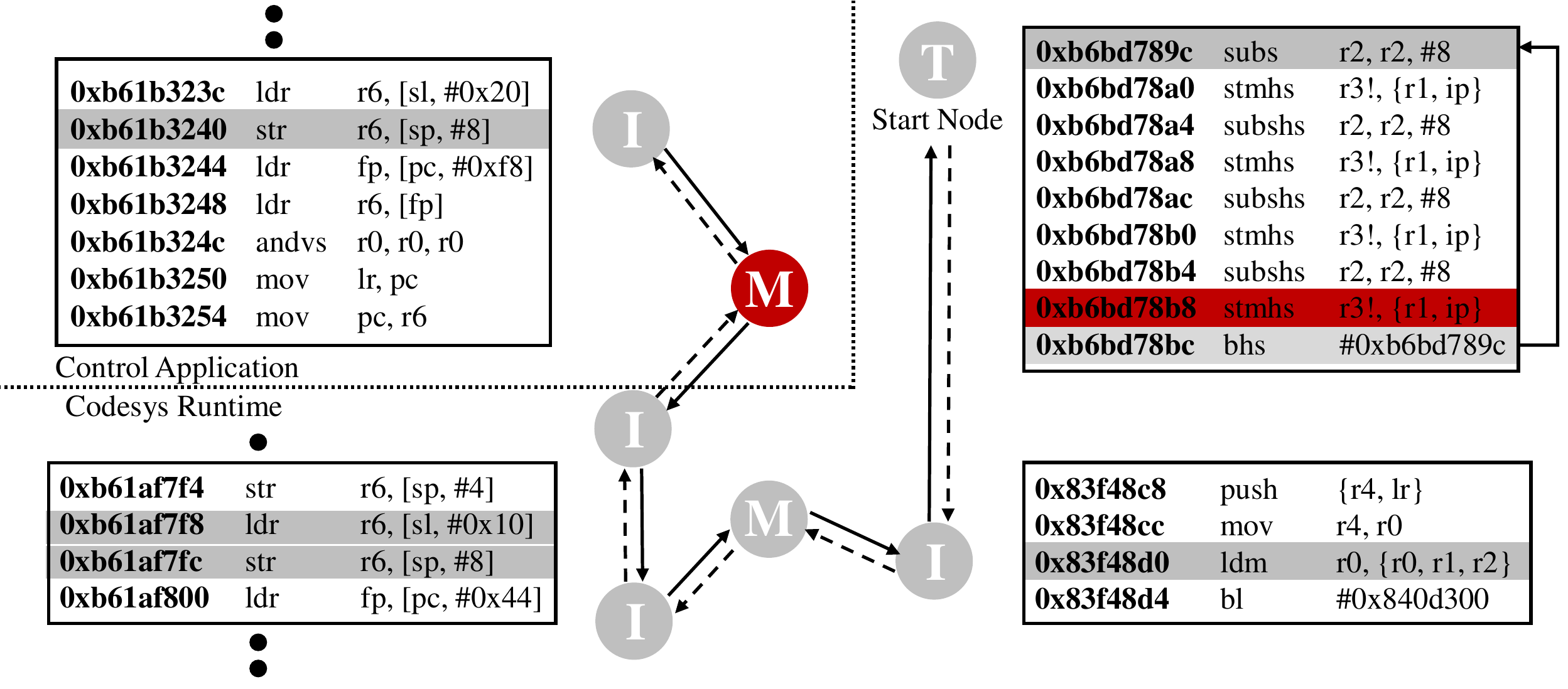}
    \small I: Instruction Node \qquad M: Memory Node \qquad T: Transition Node
    \captionof{figure}{Vulnerability localization in ICSPatch using Data Dependence Graph. The memory node (M), represented in red, shows the patching memory location of the malicious input just before the execution leaves the control application.}
    \label{fig:icspatch_vuln_localization}
    \vspace{-0.35cm}
\end{figure}

\noindent \textbf{Vulnerability Localization.} ICSPatch creates a DDG while executing the control application in~\texttt{angr} as shown in Algorithm \ref{alg:icspatch_full_mod}. In some cases, memory violation vulnerabilities in the control applications result from missing bound checks in imported functions rather than the original control logic. ICSPatch detects the exploit location, i.e., the instruction that triggers vulnerability identification. In the out-of-bounds write example shown in Figure~\ref{fig:icspatch_vuln_localization}, the \texttt{stmhs} instruction at \texttt{0xb6bd78b8} writes into the runtime stack and is therefore detected. However, this location is not the same as the patch location. Patching at such locations modifies the shared runtime function, which is also utilizable at other parts of the program. Moreover, the runtime may also call some of the available \texttt{Libc} functions instead of implementing its own; patching at this location implies hotpatching \texttt{Libc}. So, ICSPatch must identify a patching location close to the boundary between the control application and the runtime, as illustrated in Figure~\ref{fig:icspatch_vuln_localization}.

\noindent \textbf{DDG Structure and methodology.} Figure~\ref{fig:icspatch_vuln_localization} shows three main categories of DDG nodes, the instruction~\nodeParam{I}, memory~\nodeParam{M}, and transition nodes~\nodeParam{T}. Control applications utilize memory locations to pass parameters between functions, thus requiring store (\texttt{str}) and load (\texttt{ldr}) instructions, represented by instruction nodes. Store operations write to a memory location represented by memory nodes in the DDG. We need to include these nodes in our DDG to locate the input in memory that leads to the manifestation of the vulnerability (represented by~\nodeRedParam{M}). Finally, the control logic might modify the values loaded into the register or move them into a different register. Transition nodes represent such operations on the loaded memory value.

In the example displayed in Figure~\ref{fig:icspatch_vuln_localization}, the \texttt{stmhs} instruction at \texttt{0xb6bd78b8} leads to the violation of the out-of-bounds write rule. However, such bound checks are implemented by checking the value of the bound, in this case, stored in \texttt{R2}, decremented at \texttt{0xb6bd789c} with \texttt{subs} and verified at \texttt{0xb6bd78bc} with \texttt{bhs}. ICSPatch recognizes \texttt{0xb6bd789c} as the start node and traverses backward along the path represented by dashed lines reaching the instruction node at \texttt{0xb61b3240}. This \texttt{str} instruction stores the bound value at the memory location represented by~\nodeRedParam{M}, which is the value manipulated by an adversary, leading to the manifestation of the vulnerability: Out-of-bounds write in this case. Using the DDG and traversing it with its \texttt{DFS()} algorithm, ICSPatch can successfully localize vulnerabilities for control applications and select an appropriate patching location.

\noindent \textbf{Why does ICSPatch utilize DDG?} Vulnerability localization can be performed in various ways. For instance, {\scshape VulnLoc} utilizes statistics for identifying correct patching locations for 88\% of the considered 43 CVEs while considering Top-5 outputs~\cite{proc:vuln_localization}. On the other hand, the machine learning-based approach DeepVL~\cite{art:dnn_binaries_localization} achieves an accuracy of 96.9\% with low precision of 70.1\%, while Devign~\cite{art:devign} reaches accuracy of 80.24\%. ICSPatch captures memory snapshots from the development PLC, allowing simulated execution of the control application with concrete values. At the same time, control binaries typically lack optimization, so data/control flow is easier to understand/extract with precision compared to regular optimized binaries. Dynamic analysis techniques with concrete values of registers explore a single path while providing high fidelity (accurate intermediate states) and produce no false positives (every data dependency is valid)~\cite{proc:Saluki, proc:valgrind}. Since ICS devices often control and regulate physical processes in critical infrastructure, a deterministic approach with no false positives is preferable over the aforementioned probabilistic methods with false positives. Therefore, DDGs provide an accurate answer, if a vulnerability exists; if not, ICSPatch fails to patch, which is preferred to the alternative of allowing false positives, i.e., patching at incorrect locations.

\subsection{Step 2: Patch Generation \& Deployment}\label{sec:patchgen}
\noindent \textbf{Patch Generation.} In contrast to state-of-the-art embedded systems hotpatching solutions that target the RTOS~\cite{proc:hera, proc:rapid_patch}, ICSPatch focuses on a much more exotic target: the industrial control application. So, unlike state-of-the-art, it does not have an upstream source for official patches. To automate the process of hotpatching, we restrict our focus to a select category of weaknesses. In \texttt{BUILD\_PATCH()}, we create skeleton patches for all selected weaknesses and populate the patches with live values from the deployed PLC by communicating with the local patch server running on it. For instance, the out-of-bounds write/read patch compares the value of the bound with the highest permissible value to prevent illegitimate write/read to the stack. ICSPatch populates the skeleton patch with concrete values, which include the higher upper bound for the size and the live memory address of vulnerable input on the deployed PLC (shown by~\nodeRedParam{M} in Figure~\ref{fig:icspatch_vuln_localization}). On the other hand, the patch for OS command injection requires the live address of the overwritten location in the address table and the corresponding live expected value in that location, which is captured automatically from the deployed PLC.

\noindent \textbf{Patch Verification.} ICSPatch injects the hotpatch and the hook in the \texttt{angr} simulation state to rerun the scan cycle of the control application with the exploit input to check for unbounded loops, dangerous instructions (for instance, involving stack state manipulation), memory modifications outside the range of the control application stack, and the vulnerability ruleset. If the scan cycle terminates without triggering any safety checks, the patch is deemed safe for deployment. To note, ICSPatch, like other state-of-the-art solutions, cannot verify the correctness of the patch automatically, so instead, it checks for its safety~\cite{proc:rapid_patch}.

\noindent \textbf{Patch Deployment.} Patch deployment with ICSPatch requires two main components operating on the target vulnerable PLC: 1) The local patch server that communicates with the ICSPatch server over the network, which upon request from \texttt{GET\_BASE\_ADDR()} sends information such as specific base addresses, and then through \texttt{DEPLOY\_PATCH()} receives commands to verify memory location content, and write into the process memory. The local patch server also preprocesses the patch-related information received from ICSPatch before transferring it to the LKM patcher. 2) The LKM patcher is the main patching component. It executes the commands received from the ICSPatch server that are relayed via the local patch server to verify the content of a memory location or to write new content. To ensure the safety of operations, the PLC runtime must not be able to execute during the patching process. As will be discussed later in the experimental results, the critical part of patching (i.e., writing the hook) takes a negligible amount of time (ranging from 0.00022 ms to 0.00046 ms), thus running without interruption. Nevertheless, for extra caution, we introduce two more safety measures: 1) Just before the critical code, we adjust the \texttt{nice} process scheduling priority for the runtime by setting it to 19, which ensures that the runtime will only run when no other system process with higher priority wishes to execute. 2) Additionally, before entering the critical section of the patching process, our LKM temporarily disables kernel preemption and interrupt handling for the local processor using \texttt{preempt\_disable()} and \texttt{local\_irq\_disable()} respectively~\cite{book:corbet2005linux, book:love2010linux}, and re-enables them after writing the hook, ensuring an atomic operation.

At the same time, to avoid writing at inappropriate locations on the deployed PLC, our LKM verifies the memory content at said locations with the content obtained from the development PLC. Specifically, the patch and the address of the patch to be written in the address table expect empty memory locations, and only the hook that modifies the execution flow requires an exact memory content match to proceed with its installation at that specific location. The LKM patcher can write into the memory space of any process and at locations as specified by the ICSPatch server. However, patching is aborted in case of a mismatch in the memory location content during verification. The two components mentioned above operate on the deployed PLC and allow non-intrusive patching of the live control application.

\begin{figure}[t]
    \centering
    \includegraphics[width=\columnwidth, trim={0.5cm 3.2cm 7.8cm 3.1cm}, clip]{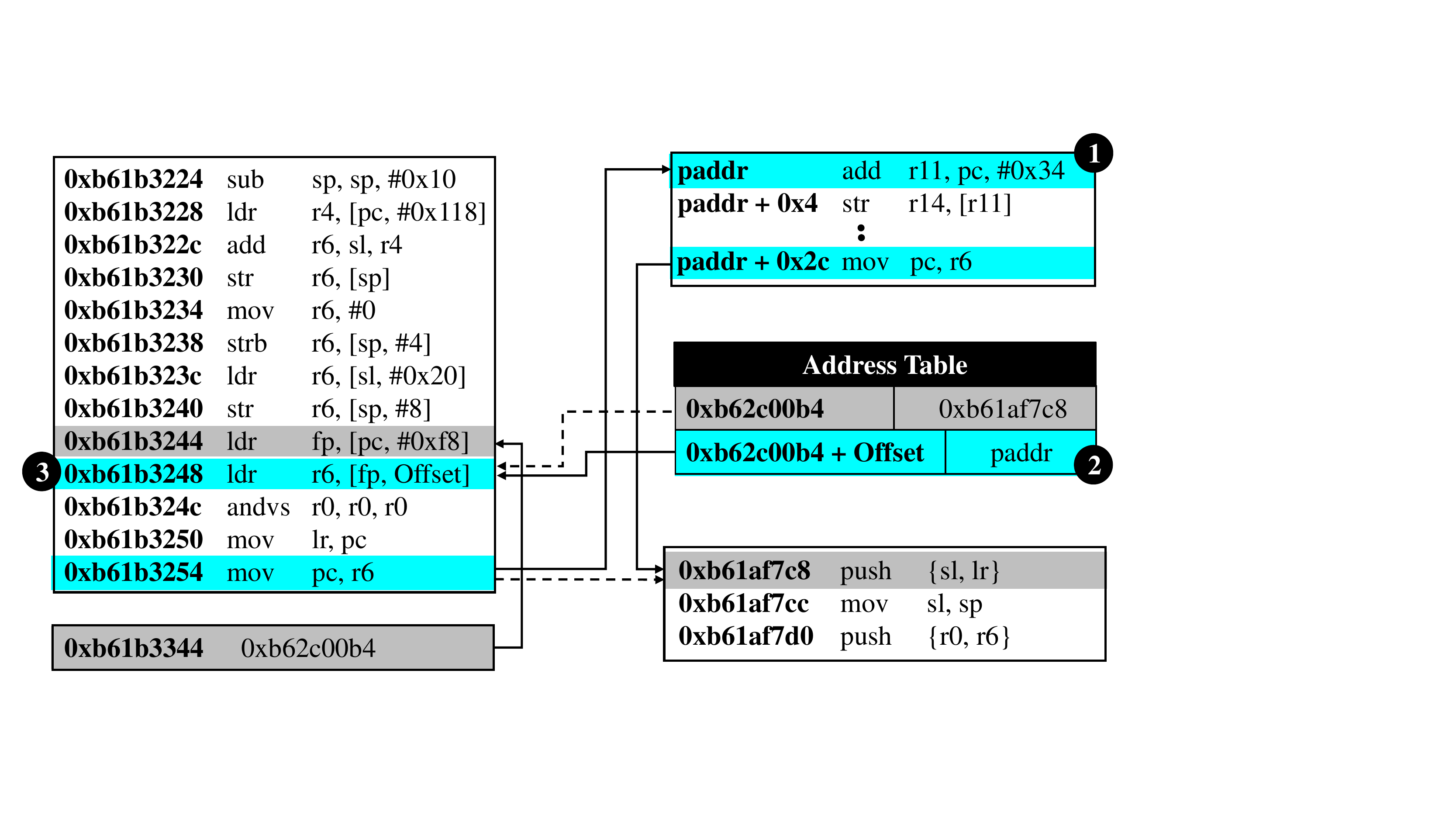}
    \captionof{figure}{Three steps and control flow modification required for patching control applications with ICSPatch.}
    \label{fig:icspatch_patch_deployment}
    \vspace{-0.35cm}
\end{figure}

The patches generated by ICSPatch require a change in the execution flow for patching the vulnerability before returning it to normal flow. As shown in Figure~\ref{fig:icspatch_patch_deployment}, three primary steps are involved. 1) ICSPatch verifies an empty memory location between the control application and the address table, writing the patch at this location (\emph{paddr}). 2) It then verifies an empty location in the address table and writes the patch address while noting its offset. To find an empty location in the address table, ICSPatch utilizes the base address of the address table loaded in an \texttt{ldr} instruction (for instance, at address \texttt{0xb61b3244} in Figure~\ref{fig:icspatch_patch_deployment}) and traverses the address table for \texttt{0x7fe} memory locations (highest value of permitted immediate offset for an \texttt{ldr} instruction) attempting to find an empty location of 32 bits. This process is performed entirely on the rehosted \texttt{angr} instance of the control application binary, not impacting the deployed PLC. 3) Finally, it modifies the \texttt{ldr} instruction at \texttt{0xb61b3248}, which was initially in this example \texttt{ldr r6, [fp]}, to \texttt{ldr r6, [fp, Offset]} for adding an offset to the original address into the address table. As a consequence of this offset, during execution, the \texttt{R6} register loads the address of the patch instead of the next function at \texttt{0xb61af7c8}. The execution flow changes to the patch, which fixes the bound value at the automatically detected memory location during vulnerability localization and passes the control back to the intended function after restoring the values of the registers employed in the patch.

\section{ICSPatch Usage Scenario}
\begin{figure}[t]
    \centering
    \includegraphics[width=0.8\columnwidth, trim={0.00cm 16.1cm 9.53cm 0.00cm}, clip]{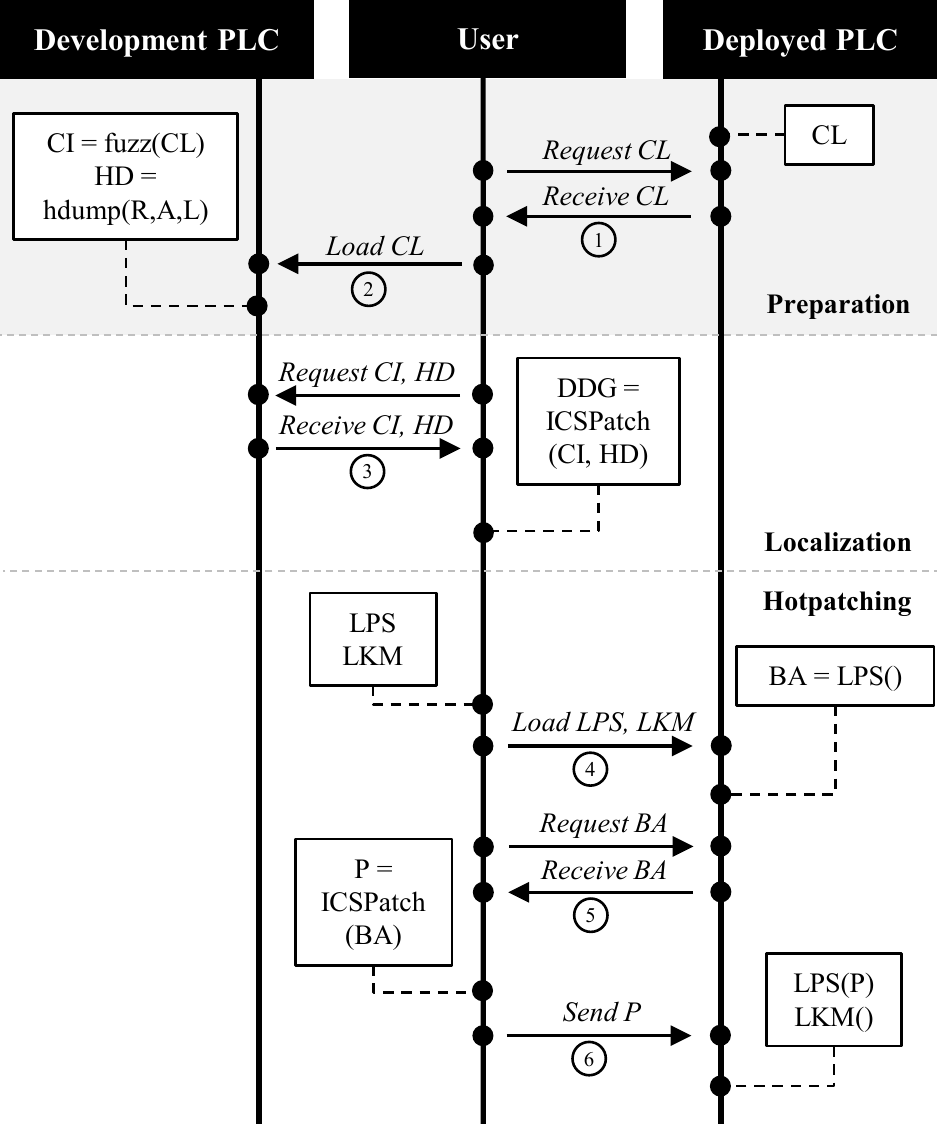}
    \newline \small CL: Control Logic \enspace CI: Crash Inputs \enspace HD: Hexdumps \enspace R: Runtime \enspace A: Application \enspace L: Libraries \enspace DDG: Data Dependency Graph \enspace LPS: Local Patch Server \enspace LKM: Loadable Kernel Module \enspace \\ BA: Base Address \enspace P: Patch
    \captionof{figure}{A chronological overview of using ICSPatch in the field. All arrows are SSH connections.}
    \label{fig:proc_graph}
    \vspace{-0.35cm}
\end{figure}

Figure~\ref{fig:proc_graph} shows three primary chronological steps involved with the real-world deployment of ICSPatch. The scenario assumes remote (SSH) access to the deployment PLC and also access to a development PLC to perform intrusive operations.

\noindent \textbf{1) Preparation (if needed).} If the user does not already have a copy of the control binary, they need to connect to the deployed PLC and extract it (\circled{1}) over the network with a simple SSH connection. Once they have the binary, they load it to a second, offline PLC (``development PLC'', \circled{2}), and a fuzzer is employed (e.g., ICSFuzz~\cite{proc:icsfuzz}) to uncover potential crash inputs in the control logic. When fuzzing yields a set of crash inputs, these are recorded, along with hexdumps of the runtime, control logic application, and memory. 

\noindent \textbf{2) Localization.} The crash files described before are retrieved by the user \circled{3}, and ICSPatch is executed, localizing the vulnerability using DDGs.

\noindent \textbf{3) Hotpatching.} To hotpatch the vulnerability, the local patch server and LKM are loaded onto the deployed PLC \circled{4}. The local patch server extracts the runtime base address \circled{5}, which is provided to ICSPatch to create a patch. Consequently, the user sends the patch to the deployed PLC \circled{6}, where the local patch server interfaces with the LKM to patch the control application in memory.

\section{Dataset and Vulnerabilities} \label{sec:iec_vulnerabilities}
Table~\ref{table:cwe_vuln} presents the top 5 software weaknesses for 2021~\cite{misc:mitre_cwe}. We observe that out-of-bounds write, read, improper input validation, and OS command injection apply to control binaries and can affect PLCs. Cross-site scripting, on the other hand, is not an applicable threat to the control application and, therefore, is not considered. Software weaknesses can affect PLCs since the control binary runs in the runtime context, inherits its privileges, and shares its memory spaces. 

To evaluate the ability of ICSPatch to localize and patch vulnerabilities, we created a synthetic dataset of control application binaries from five critical infrastructure sectors, including aircraft flight control, anaerobic digestion reactor, smart grid, chemical, and desalination plants, with a total of 24 vulnerable binaries. Table~\ref{table:iec_dataset} provides a summary of the synthetic dataset, including the shared library, the imported functions utilized for each example control application, and the corresponding critical infrastructure.

\begin{table}
\centering
\caption{Top 5 Common Weakness Enumeration (CWEs) software weaknesses for 2021 as reported by MITRE~\cite{misc:mitre_cwe}.}
\setlength\tabcolsep{20pt}
\label{table:cwe_vuln}
\resizebox{\columnwidth}{!}{

\begin{tabular}{clc} 
\toprule
\textbf{Rank} & \multicolumn{1}{c}{\textbf{CWE}} & \textbf{Description}       \\ 
\midrule
1             & CWE-787                          & Out-of-bounds write        \\
2             & CWE-79                           & Cross-site scripting       \\
3             & CWE-125                          & Out-of-bounds read         \\
4             & CWE-20                           & Improper input validation  \\
5             & CWE-78                           & OS command injection       \\
\bottomrule
\end{tabular}
}
\end{table}

\begin{figure}[t]
    \centering
    \includegraphics[width=0.9\columnwidth, trim={11.7cm 1.8cm 1.6cm 3cm}, clip]{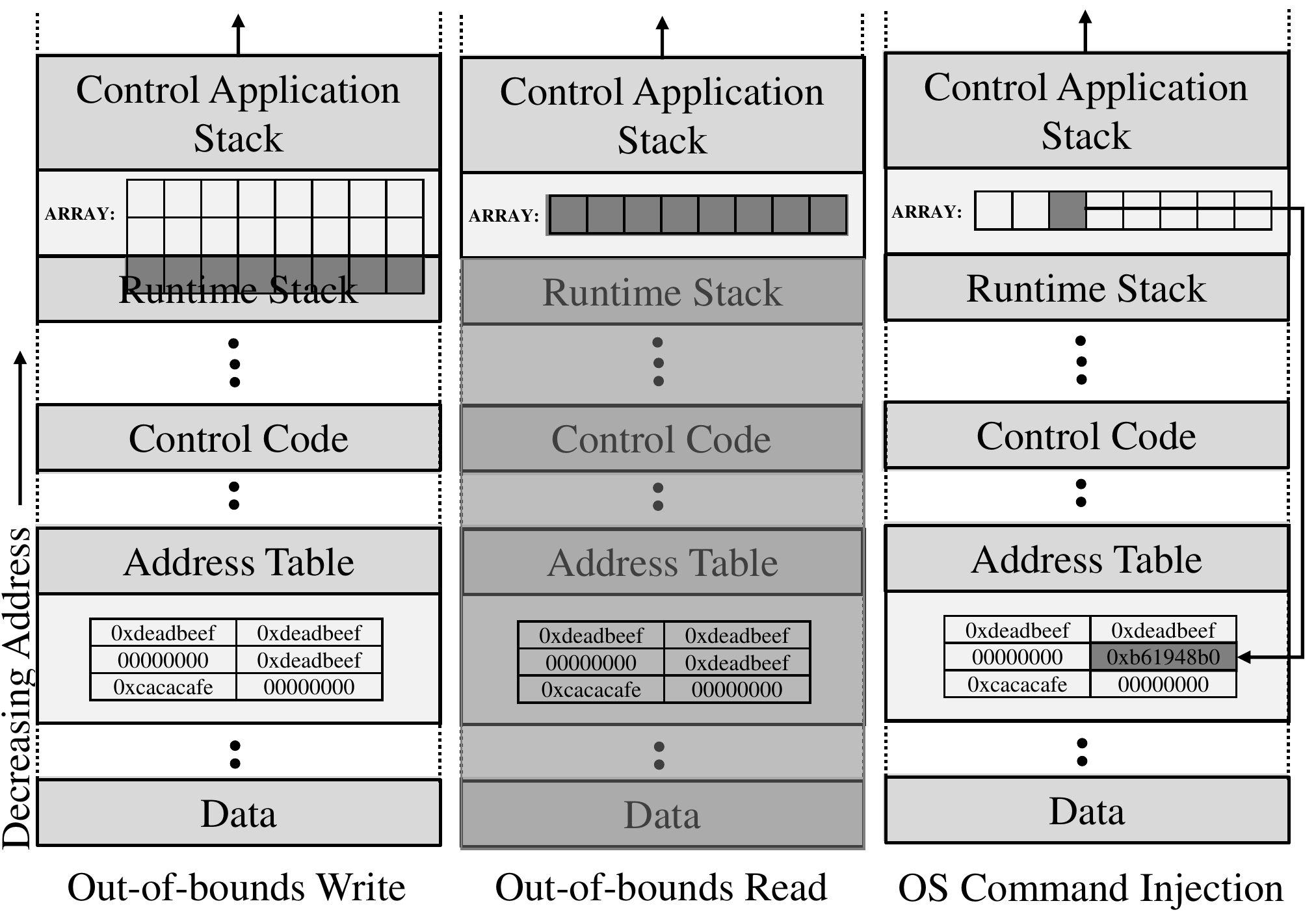}
    \caption{Typical operation and the manifestation of out-of-bounds write, read, and OS command injection vulnerabilities in PLCs. Here, the control code represents the currently executing control application.}
    \label{fig:stack_iec_vulnerabilities}
    \vspace{-0.35cm}
\end{figure}

\noindent \textbf{CWE-787 Out-of-bounds write.} An adversary can manipulate the size of the destination array by setting it to a size smaller than that of the source array in functions such as \texttt{SysMemSet}, \texttt{SysMemMove}, \texttt{MemCpy} and \texttt{BitCpy}, overwriting memory locations on the stack. These overwritten memory locations can be confined to the stack of the control application or can also cross over to the stack region of the runtime functions, crashing the \texttt{MainTask}. As a result, the control application or the runtime might crash (will not execute) due to overwritten variables or return addresses, as shown in Figure~\ref{fig:stack_iec_vulnerabilities}.

\begin{table}[t]
\centering
\caption{A diverse synthetic control application dataset for testing ICSPatch.}
\label{table:iec_dataset}
\resizebox{\columnwidth}{!}{

\begin{tabular}{cccccccccccccccccccccc} 
    \toprule
    \multirow{2}{*}{\begin{tabular}[c]{@{}c@{}}\textbf{Shared}\\\textbf{ Library}\end{tabular}} & \multirow{2}{*}{\begin{tabular}[c]{@{}c@{}}\textbf{Imported}\\\textbf{ Functions}\end{tabular}} & \multicolumn{4}{c}{\begin{tabular}[c]{@{}c@{}}\textbf{Aircraft Flight}\\\textbf{ Control}\end{tabular}}                                                                                 & \multicolumn{4}{c}{\begin{tabular}[c]{@{}c@{}}\textbf{Anaerobic Dig}\\\textbf{ estion Reactor}\end{tabular}}                                                                            & \multicolumn{4}{c}{\begin{tabular}[c]{@{}c@{}}\textbf{Chemical}\\\textbf{ Plant}\end{tabular}}                                                                                          & \multicolumn{4}{c}{\begin{tabular}[c]{@{}c@{}}\textbf{Desalination}\\\textbf{ Plant}\end{tabular}}                                                                                      & \multicolumn{4}{c}{\begin{tabular}[c]{@{}c@{}}\textbf{Smart}\\\textbf{ Grid}\end{tabular}}                                                                                               \\ 
    \cmidrule(l){3-22}
                                                                                                &                                                                                                 & \rotatebox[origin=c]{90}{\textbf{CWE-787}} & \rotatebox[origin=c]{90}{\textbf{CWE-125}} & \rotatebox[origin=c]{90}{\textbf{CWE-78}} & \rotatebox[origin=c]{90}{\textbf{CWE-20}} & \rotatebox[origin=c]{90}{\textbf{CWE-787}} & \rotatebox[origin=c]{90}{\textbf{CWE-125}} & \rotatebox[origin=c]{90}{\textbf{CWE-78}} & \rotatebox[origin=c]{90}{\textbf{CWE-20}} & \rotatebox[origin=c]{90}{\textbf{CWE-787}} & \rotatebox[origin=c]{90}{\textbf{CWE-125}} & \rotatebox[origin=c]{90}{\textbf{CWE-78}} & \rotatebox[origin=c]{90}{\textbf{CWE-20}} & \rotatebox[origin=c]{90}{\textbf{CWE-787}} & \rotatebox[origin=c]{90}{\textbf{CWE-125}} & \rotatebox[origin=c]{90}{\textbf{CWE-78}} & \rotatebox[origin=c]{90}{\textbf{CWE-20}} & \rotatebox[origin=c]{90}{\textbf{CWE-787}} & \rotatebox[origin=c]{90}{\textbf{CWE-125}} & \rotatebox[origin=c]{90}{\textbf{CWE-78}} & \rotatebox[origin=c]{90}{\textbf{CWE-20}}  \\ 
    \midrule
    \multirow{2}{*}{SysMem23}                                                                   & SysMemSet                                                                                       & \CIRCLE                                    & \Circle                                    & \Circle                                   & \Circle                                   & \Circle                                    & \Circle                                    & \Circle                                   & \Circle                                   & \Circle                                    & \Circle                                    & \Circle                                   & \Circle                                   & \Circle                                    & \Circle                                    & \Circle                                   & \Circle                                   & \Circle                                    & \Circle                                    & \Circle                                   & \Circle                                    \\
                                                                                                & SysMemMove                                                                                      & \Circle                                    & \Circle                                    & \Circle                                   & \Circle                                   & \CIRCLE                                    & \Circle                                    & \Circle                                   & \Circle                                   & \Circle                                    & \Circle                                    & \Circle                                   & \Circle                                   & \Circle                                    & \Circle                                    & \Circle                                   & \Circle                                   & \Circle                                    & \Circle                                    & \Circle                                   & \Circle                                    \\ 
    \midrule
    \multirow{3}{*}{SysMem}                                                                     & SysMemSet                                                                                       & \Circle                                    & \Circle                                    & \Circle                                   & \Circle                                   & \Circle                                    & \Circle                                    & \Circle                                   & \Circle                                   & \Circle                                    & \Circle                                    & \Circle                                   & \Circle                                   & \CIRCLE                                    & \Circle                                    & \Circle                                   & \Circle                                   & \Circle                                    & \Circle                                    & \Circle                                   & \CIRCLE                                    \\
                                                                                                & SysMemMove                                                                                      & \Circle                                    & \Circle                                    & \Circle                                   & \Circle                                   & \Circle                                    & \Circle                                    & \Circle                                   & \Circle                                   & \Circle                                    & \CIRCLE                                    & \Circle                                   & \Circle                                   & \Circle                                    & \Circle                                    & \Circle                                   & \CIRCLE                                   & \Circle                                    & \Circle                                    & \Circle                                   & \Circle                                    \\
                                                                                                & SysMemCpy                                                                                       & \Circle                                    & \Circle                                    & \Circle                                   & \Circle                                   & \Circle                                    & \Circle                                    & \Circle                                   & \Circle                                   & \Circle                                    & \Circle                                    & \Circle                                   & \CIRCLE                                   & \Circle                                    & \CIRCLE                                    & \Circle                                   & \Circle                                   & \Circle                                    & \Circle                                    & \Circle                                   & \Circle                                    \\ 
    \midrule
    \multirow{3}{*}{MemUtils}                                                                   & MemSet                                                                                          & \Circle                                    & \Circle                                    & \Circle                                   & \CIRCLE                                   & \CIRCLE                                    & \Circle                                    & \Circle                                   & \Circle                                   & \Circle                                    & \Circle                                    & \Circle                                   & \Circle                                   & \Circle                                    & \Circle                                    & \Circle                                   & \Circle                                   & \Circle                                    & \Circle                                    & \Circle                                   & \Circle                                    \\
                                                                                                & MemCpy                                                                                          & \Circle                                    & \CIRCLE                                    & \Circle                                   & \Circle                                   & \CIRCLE                                    & \CIRCLE                                    & \Circle                                   & \Circle                                   & \CIRCLE                                    & \Circle                                    & \Circle                                   & \Circle                                   & \Circle                                    & \Circle                                    & \Circle                                   & \Circle                                   & \CIRCLE                                    & \Circle                                    & \Circle                                   & \Circle                                    \\
                                                                                                & BitCpy                                                                                          & \CIRCLE                                    & \Circle                                    & \Circle                                   & \Circle                                   & \Circle                                    & \CIRCLE                                    & \Circle                                   & \CIRCLE                                   & \Circle                                    & \Circle                                    & \Circle                                   & \Circle                                   & \Circle                                    & \Circle                                    & \Circle                                   & \Circle                                   & \Circle                                    & \CIRCLE                                    & \Circle                                   & \Circle                                    \\ 
    \hline\hline
    \multicolumn{2}{c}{\textbf{IEC 61131-3 code}}                                                                                                                                                   & \multicolumn{4}{c}{}                                                                                 & \multicolumn{4}{c}{}                                                                            & \multicolumn{4}{c}{}                                                                                          & \multicolumn{4}{c}{}                                                                                      & \multicolumn{4}{c}{}                                                                                               \\ 
    \hline
    \multicolumn{2}{c}{Out-of-bounds array index}                                                                                                                                       & \Circle                                    & \Circle                                    & \CIRCLE                                   & \Circle                                   & \Circle                                    & \Circle                                    & \CIRCLE                                   & \Circle                                   & \Circle                                    & \Circle                                    & \CIRCLE                                   & \Circle                                   & \Circle                                    & \Circle                                    & \CIRCLE                                   & \Circle                                   & \Circle                                    & \Circle                                    & \CIRCLE                                   & \Circle                                    \\
    \bottomrule
\end{tabular}

}
\begin{tablenotes}
    \centering
    \scriptsize
    \item CWE-787/CWE-125: Out-of-Bounds Write/Read \enspace CWE-78: OS Command Injection 
    \item CWE-20: Improper Input Validation
\end{tablenotes}
\vspace{-0.35cm}
\end{table}

\noindent \textbf{CWE-125 Out-of-bounds read.} These vulnerabilities can manifest in functions missing bound checks when reading from a larger destination array compared to the source, enabling an adversary to read from any of the selected memory regions, including the stack, the code and data sections of the runtime, and the control application as shown in Figure~\ref{fig:stack_iec_vulnerabilities}. Out-of-bounds read samples do not lead to runtime or control application crashes.

\noindent \textbf{CWE-78 OS command injection.} In these examples, an adversary modifies the execution flow to execute the injected payload or point to the beginning of a ROP chain. The examples achieve this by using an array and a malicious index manipulated by the adversary to modify the address of the following function in the address table to redirect the control flow to the malicious code rather than the expected function, as shown in Figure~\ref{fig:stack_iec_vulnerabilities}.

\noindent \textbf{CWE-20 Improper input validation.} An input to the control application can be from the connected sensors or other industrial communication protocols such as Open Industry Standard Unified Architecture (OPC UA). An adversary can manipulate inputs sent via the network during transit, requiring validation checks in the control application. Abstractly, all the other presented weaknesses could result from improper input validation, and we have already explored them individually. Therefore, for our dataset, we enrich it with more out-of-bounds write weaknesses.

\section{Performance Evaluation}
To evaluate the performance of ICSPatch, we test it against the 24 control applications presented in Section~\ref{sec:iec_vulnerabilities}. We use two PLCs with different computational capabilities: A Wago PFC100 with a single-core Cortex A8 processor operating at 600 MHz and a Wago PFC200 operating at 1 GHz.

\begin{table*}[t]
\centering
\caption{ICSPatch execution timings and overheads for the 24 vulnerable binaries.}
\label{table:latency}
\setlength\tabcolsep{7pt}
\resizebox{\linewidth}{!}{
\begin{tabular}{clcccccccccc}
\hline
\multirow{3}{*}{\textbf{\begin{tabular}[c]{@{}c@{}}Critical\\ Infrastructure\end{tabular}}}   & \multicolumn{1}{c}{\multirow{3}{*}{\textbf{\begin{tabular}[c]{@{}c@{}}Vulnera-\\ -bility\end{tabular}}}} & \multicolumn{5}{c}{\textbf{Time (ms)}}                                                                                                                                                                                                                                                                                                                                         & \multicolumn{3}{c}{\textbf{Mean Execution Time ($\mu$s)}}                                                         & \multirow{3}{*}{\textbf{\begin{tabular}[c]{@{}c@{}}Achieved\\ Scan Cycle\\ ($\mu$s)\end{tabular}}} & \multirow{3}{*}{\textbf{\begin{tabular}[c]{@{}c@{}}Memory\\ (Bytes)\end{tabular}}} \\ \cline{3-10}
                                                                                              & \multicolumn{1}{c}{}                                                                                     & \multirow{2}{*}{\textbf{\begin{tabular}[c]{@{}c@{}}Vulnerability\\ Localization\end{tabular}}} & \multirow{2}{*}{\textbf{\begin{tabular}[c]{@{}c@{}}Patch\\ Generation\end{tabular}}} & \multirow{2}{*}{\textbf{\begin{tabular}[c]{@{}c@{}}Patch\\ Verification\\ (s)\end{tabular}}} & \multicolumn{2}{c}{\textbf{\begin{tabular}[c]{@{}c@{}}Patch\\ Deployment\end{tabular}}} & \multirow{2}{*}{\textbf{Pre-patch}} & \multirow{2}{*}{\textbf{Post-patch}} & \multirow{2}{*}{\textbf{Difference}} &                                                                                                    &                                                                                    \\ \cline{6-7}
                                                                                              & \multicolumn{1}{c}{}                                                                                     &                                                                                                &                                                                                      &                                                                                              & \textbf{Total}  & \textbf{\begin{tabular}[c]{@{}c@{}}Critical\\ ($\mu$s)\end{tabular}}  &                                     &                                      &                                      &                                                                                                    &                                                                                    \\ \hline
\multirow{5}{*}{\textbf{\begin{tabular}[c]{@{}c@{}}Aircraft\\ Flight\\ Control\end{tabular}}} & CWE-20                                                                                                   & 3.06                                                                                           & 178.57                                                                               & 7.71                                                                                         & 252.13          & 0.22                                                                  & 20.09                               & 21.06                                & 0.97                                 & 69.59                                                                                              & 64                                                                                 \\
                                                                                              & CWE-787                                                                                                  & 6.72                                                                                           & 166.39                                                                               & 67.18                                                                                        & 332.73          & 0.3                                                                   & 18.05                               & 20.04                                & 1.99                                 & 77.2                                                                                               & 64                                                                                 \\
                                                                                              & CWE-787                                                                                                  & 4.42                                                                                           & 143.68                                                                               & 33.43                                                                                        & 232.2           & 0.26                                                                  & 25.4                                & 27.19                                & 1.8                                  & 74.94                                                                                              & 64                                                                                 \\
                                                                                              & CWE-125                                                                                                  & 4.85                                                                                           & 178.59                                                                               & 28.93                                                                                        & 252.24          & 0.46                                                                  & 21.6                                & 21.78                                & 0.17                                 & 74.54                                                                                              & 64                                                                                 \\
                                                                                              & CWE-78                                                                                                   & 1.54                                                                                           & 203.03                                                                               & 9                                                                                            & 230.17          & 0.3                                                                   & 16.61                               & 20.5                                 & 3.89                                 & 72.87                                                                                              & 56                                                                                 \\ \hline
\multirow{7}{*}{\textbf{\begin{tabular}[c]{@{}c@{}}Anaerobic\\ Reactor\end{tabular}}}         & CWE-20                                                                                                   & 5.05                                                                                           & 134.98                                                                               & 10                                                                                           & 234.42          & 0.31                                                                  & 20.04                               & 21.04                                & 0.99                                 & 134.46                                                                                             & 64                                                                                 \\
                                                                                              & CWE-787                                                                                                  & 3.78                                                                                           & 126.87                                                                               & 2.86                                                                                         & 232.07          & 0.35                                                                  & 17.02                               & 17.07                                & 0.05                                 & 71.46                                                                                              & 64                                                                                 \\
                                                                                              & CWE-787                                                                                                  & 4.82                                                                                           & 130.18                                                                               & 7.17                                                                                         & 246.06          & 0.24                                                                  & 19.99                               & 20.24                                & 0.25                                 & 147.74                                                                                             & 64                                                                                 \\
                                                                                              & CWE-787                                                                                                  & 4.44                                                                                           & 124.95                                                                               & 2.16                                                                                         & 223.7           & 0.22                                                                  & 21.32                               & 21.37                                & 0.05                                 & 74.49                                                                                              & 64                                                                                 \\
                                                                                              & CWE-125                                                                                                  & 5.91                                                                                           & 125.24                                                                               & 4.62                                                                                         & 234.49          & 0.23                                                                  & 16.89                               & 18.94                                & 2.04                                 & 77.3                                                                                               & 64                                                                                 \\
                                                                                              & CWE-125                                                                                                  & 4.96                                                                                           & 221.38                                                                               & 169.1                                                                                        & 236.26          & 0.28                                                                  & 23.98                               & 28.08                                & 4.09                                 & 152.96                                                                                             & 64                                                                                 \\
                                                                                              & CWE-78                                                                                                   & 1.44                                                                                           & 171.11                                                                               & 6.16                                                                                         & 298.04          & 0.3                                                                   & 15.10                               & 19.3                                 & 4.19                                 & 71.7                                                                                               & 56                                                                                 \\ \hline
\multirow{4}{*}{\textbf{\begin{tabular}[c]{@{}c@{}}Chemical\\ Plant\end{tabular}}}            & CWE-20                                                                                                   & 5                                                                                              & 126                                                                                  & 1.85                                                                                         & 254.16          & 0.3                                                                   & 14.82                               & 17.45                                & 2.63                                 & 67.75                                                                                              & 64                                                                                 \\
                                                                                              & CWE-787                                                                                                  & 3.76                                                                                           & 183.18                                                                               & 424.23                                                                                       & 236.42          & 0.31                                                                  & 36.62                               & 54.03                                & 17.4                                 & 148.54                                                                                             & 64                                                                                 \\
                                                                                              & CWE-125                                                                                                  & 6.61                                                                                           & 170.95                                                                               & 17.83                                                                                        & 253.94          & 0.23                                                                  & 21.37                               & 25.8                                 & 4.42                                 & 71.41                                                                                              & 64                                                                                 \\
                                                                                              & CWE-78                                                                                                   & 1.64                                                                                           & 127.89                                                                               & 20.26                                                                                        & 252.72          & 0.26                                                                  & 24.09                               & 26.64                                & 2.54                                 & 72.38                                                                                              & 56                                                                                 \\ \hline
\multirow{4}{*}{\textbf{\begin{tabular}[c]{@{}c@{}}Desalination\\ Plant\end{tabular}}}        & CWE-20                                                                                                   & 5.44                                                                                           & 134.94                                                                               & 7.4                                                                                          & 244             & 0.24                                                                  & 18.6348                             & 19.88                                & 1.25                                 & 75.82                                                                                              & 64                                                                                 \\
                                                                                              & CWE-787                                                                                                  & 4.81                                                                                           & 127.11                                                                               & 3.32                                                                                         & 238.35          & 0.23                                                                  & 15.717                              & 18.07                                & 2.35                                 & 75.79                                                                                              & 64                                                                                 \\
                                                                                              & CWE-125                                                                                                  & 4.94                                                                                           & 139.36                                                                               & 11.25                                                                                        & 241.02          & 0.25                                                                  & 19.3252                             & 19.99                                & 0.66                                 & 73.08                                                                                              & 64                                                                                 \\
                                                                                              & CWE-78                                                                                                   & 1.52                                                                                           & 133.81                                                                               & 5.5                                                                                          & 230.13          & 0.26                                                                  & 17.483                              & 20.52                                & 3.04                                 & 80.5                                                                                               & 56                                                                                 \\ \hline
\multirow{4}{*}{\textbf{\begin{tabular}[c]{@{}c@{}}Smart\\ Grid\end{tabular}}}                & CWE-20                                                                                                   & 3.95                                                                                           & 133.9                                                                                & 3                                                                                            & 264.83          & 0.27                                                                  & 15.1286                             & 17.15                                & 2.02                                 & 65.71                                                                                              & 64                                                                                 \\
                                                                                              & CWE-787                                                                                                  & 3.64                                                                                           & 134.14                                                                               & 9.2                                                                                          & 247.91          & 0.22                                                                  & 26.0833                             & 27.02                                & 0.94                                 & 83.69                                                                                              & 64                                                                                 \\
                                                                                              & CWE-125                                                                                                  & 5.73                                                                                           & 126.27                                                                               & 4.6                                                                                          & 227.34          & 0.23                                                                  & 20.0931                             & 22.98                                & 2.89                                 & 97.48                                                                                              & 64                                                                                 \\
                                                                                              & CWE-78                                                                                                   & 1.46                                                                                           & 222.22                                                                               & 6.6                                                                                          & 232.86          & 0.34                                                                  & 25.2406                             & 27.31                                & 2.07                                 & 77.12                                                                                              & 56                                                                                 \\ \hline
\end{tabular}
}
\begin{tablenotes}
    \centering
    \scriptsize
    \item CWE-20: Improper Input Validation \qquad CWE-787: Out-of-Bounds Write \qquad CWE-125: Out-of-Bounds Read \qquad CWE-78: OS Command Injection
\end{tablenotes}
\end{table*}

\noindent{\bf Vulnerability Localization Accuracy.} To test the accuracy of ICSPatch, we utilize our synthetic application binary dataset with 24 binaries and the 20 vulnerable binaries from the ICSFuzz dataset~\cite{proc:icsfuzz}. We loaded these control applications on Wago PFC 100 and successfully detected all the vulnerabilities in the 44 vulnerable binaries (100\% localization accuracy). However, it should be noted that ICSPatch detects vulnerabilities based on specific violation rules, and any function that does not follow this pattern might go undetected. The ICSFuzz dataset consisted primarily of out-of-bounds write vulnerability manifesting due to various imported functions and, in some examples, incorrect index to an array, enabling successful detection.

\begin{table*}[t]
\centering
\caption{Detailed breakdown of ICSPatch used on the Aircraft Flight Control CWE-20 vulnerable binary.}
\label{table:detail_time}
\setlength\tabcolsep{8pt}
\resizebox{\linewidth}{!}{
\begin{tabular}{c|c|ccc|cccc|cccccc}
\hline
\textbf{Phases}                 & \textbf{Preparation}                                                          & \multicolumn{3}{c|}{\textbf{Vulnerability Localization}}                                                                                                                                                                                                                         & \multicolumn{4}{c|}{\textbf{Patch Generation}}                                                                                                                                                                                                                                                                                                                                         & \multicolumn{6}{c}{\textbf{Patch Deployment}}                                                                                           \\ \hline
\multirow{2}{*}{\textbf{Steps}} & \multirow{2}{*}{\begin{tabular}[c]{@{}c@{}}Hexdump\\ Extraction\end{tabular}} & \multicolumn{1}{c|}{\multirow{2}{*}{\begin{tabular}[c]{@{}c@{}}Load\\ Hexdumps\end{tabular}}} & \multicolumn{1}{c|}{\multirow{2}{*}{\begin{tabular}[c]{@{}c@{}}Control App\\ Execution\end{tabular}}} & \multirow{2}{*}{\begin{tabular}[c]{@{}c@{}}DDG\\ Traversal\end{tabular}} & \multicolumn{1}{c|}{\multirow{2}{*}{\begin{tabular}[c]{@{}c@{}}Locate Live\\ Addresses\end{tabular}}} & \multicolumn{1}{c|}{\multirow{2}{*}{\begin{tabular}[c]{@{}c@{}}Hook\\ Creation\end{tabular}}} & \multicolumn{1}{c|}{\multirow{2}{*}{\begin{tabular}[c]{@{}c@{}}Patch\\ Creation\end{tabular}}} & \multirow{2}{*}{\begin{tabular}[c]{@{}c@{}}Patch\\ Verification\end{tabular}} & \multicolumn{1}{c|}{MV} & \multicolumn{1}{c|}{MW} & \multicolumn{1}{c|}{MV} & \multicolumn{1}{c|}{MW} & \multicolumn{1}{c|}{MV} & MW    \\ \cline{10-15} 
                                &                                                                               & \multicolumn{1}{c|}{}                                                                         & \multicolumn{1}{c|}{}                                                                                 &                                                                          & \multicolumn{1}{c|}{}                                                                                 & \multicolumn{1}{c|}{}                                                                         & \multicolumn{1}{c|}{}                                                                          &                                                                               & \multicolumn{2}{c|}{Address Table}                & \multicolumn{2}{c|}{Patch}                        & \multicolumn{2}{c}{Hook}        \\ \hline
\textbf{Device}                 & \begin{tabular}[c]{@{}c@{}}Development\\ PLC\end{tabular}                     & \multicolumn{3}{c|}{ICSPatch (\texttt{angr})}                                                                                                                                                                                                                                    & \multicolumn{1}{c|}{\begin{tabular}[c]{@{}c@{}}Deployed\\ PLC\end{tabular}}                           & \multicolumn{3}{c|}{ICSPatch}                                                                                                                                                                                                                                                  & \multicolumn{6}{c}{Deployed PLC}                                                                                                        \\ \hline
\textbf{Time (s)}               & 733.11                                                                        & 52.02                                                                                         & 4.73                                                                                                  & 0.003                                                                    & 0.06                                                                                                  & 0.09                                                                                          & 0.03                                                                                           & 7.71                                                                          & 0.016                   & 0.033                   & 0.054                   & 0.055                   & 0.05                    & 0.043 \\ \hline
\end{tabular}
}
\begin{tablenotes}
    \centering
    \scriptsize
    \item MV: Memory Verification \qquad MW: Memory Write 
\end{tablenotes}
\vspace{-0.15in}
\end{table*}

\noindent{\bf ICSPatch Execution Timings.} Table~\ref{table:latency} shows that ICSPatch can successfully localize vulnerabilities in the control application in under a minute for most cases, except for the out-of-bounds write example for the Chemical Plant infrastructure, mainly due to the execution time of the control application in \texttt{angr} (310.4 s) because of a large number of loops in the control logic. At the same time, patch generation with ICSPatch ranges from $\approx$124 ms to $\approx$222 ms, while patch deployment ranges from $\approx$223 ms to $\approx$332 ms. This timing includes everything, such as empty memory location verification, writing the patch in memory, verifying the hook location, and the hook writing time. The last step, specifically the LKM part involving hook injection, is the critical step (since it modifies the control flow of the ICS binary) and appears in Table~\ref{table:latency} as the ``Critical'' column. As evident in the Critical column, LKM hook writing time is extremely fast, in the order of \textmu s, significantly reducing the probability of the runtime executing while hook writing takes place. Nevertheless, ICSPatch takes additional precautions for ensuring atomic hook writing operation as outlined in Section \ref{sec:patchgen}. Table~\ref{table:detail_time} takes one of the binaries (CWE-20 for Aircraft Flight Control) and further breaks down the timings. The time for writing the patch hook in memory, as shown in Table~\ref{table:detail_time}, includes the ICSPatch communication with the user system, the local patch server time, and the LKM time. Table~\ref{table:latency} presents only the critical LKM time, which is a subset of the total time.

\noindent{\bf ICSPatch Overhead.} Table~\ref{table:latency} also shows the latency incurred by the control application execution as a result of the patch in \textmu s. The increase in the execution time is negligible (0.05-17.4 \textmu s) compared to typical scan cycles of critical infrastructures, which range from a few milliseconds to a few seconds (exact scan cycle values are process and setup specific), further illustrated in the case study in Section~\ref{sec:case_study}. Patching increases latency by only a few \textmu s; however, the out-of-bounds write example for the chemical plant is an exception with an increase of $\approx$17.4 \textmu s. This increase is independent of the patch and is a consequence of the structure of the control application. Unlike other examples, an exploit input, in this case, triggers the execution of additional \texttt{for} loops for larger array initialization. The patch only fixes the input to the \texttt{MemUtils.MemCpy} that results in the out-of-bounds write. Initializing a larger array due to an exploit input is a legitimate operation as it does not overwrite any memory location, leading to an increase in time. 

With regards to the extra memory needed for the patch, presented in Table~\ref{table:latency}, there are three crucial parts to the patches generated by ICSPatch: The main patch, the address of the patch written in an empty location inside the address table, and the hook. The patch address is 32 bits for our target platforms, and the hook size depends on the target compilation mode for the runtime. In our experiments, the hook for our Wago PFC100 (ARM) is 32 bits, as it consists of modifying an \texttt{ldr} instruction with an immediate operand. The main patch for out-of-bounds write and read vulnerabilities is based on 13 assembly code lines and the 32 bit bound limit. Combined with the patch address and the hook, this results in a 64-byte patch. On the other hand, the skeleton patch for resolving OS command injection vulnerability does not include a bound limit and also does not require us to load the base address to branch to the patch location, thus saving an additional 4 bytes by the omission of an \texttt{ldr} instruction, making the total patch size 56 bytes.

\begin{figure}[t]
\minipage{0.485\columnwidth}
  \includegraphics[width=\linewidth, trim={0.2cm 0.2cm 0.1cm 0.2cm}, clip]{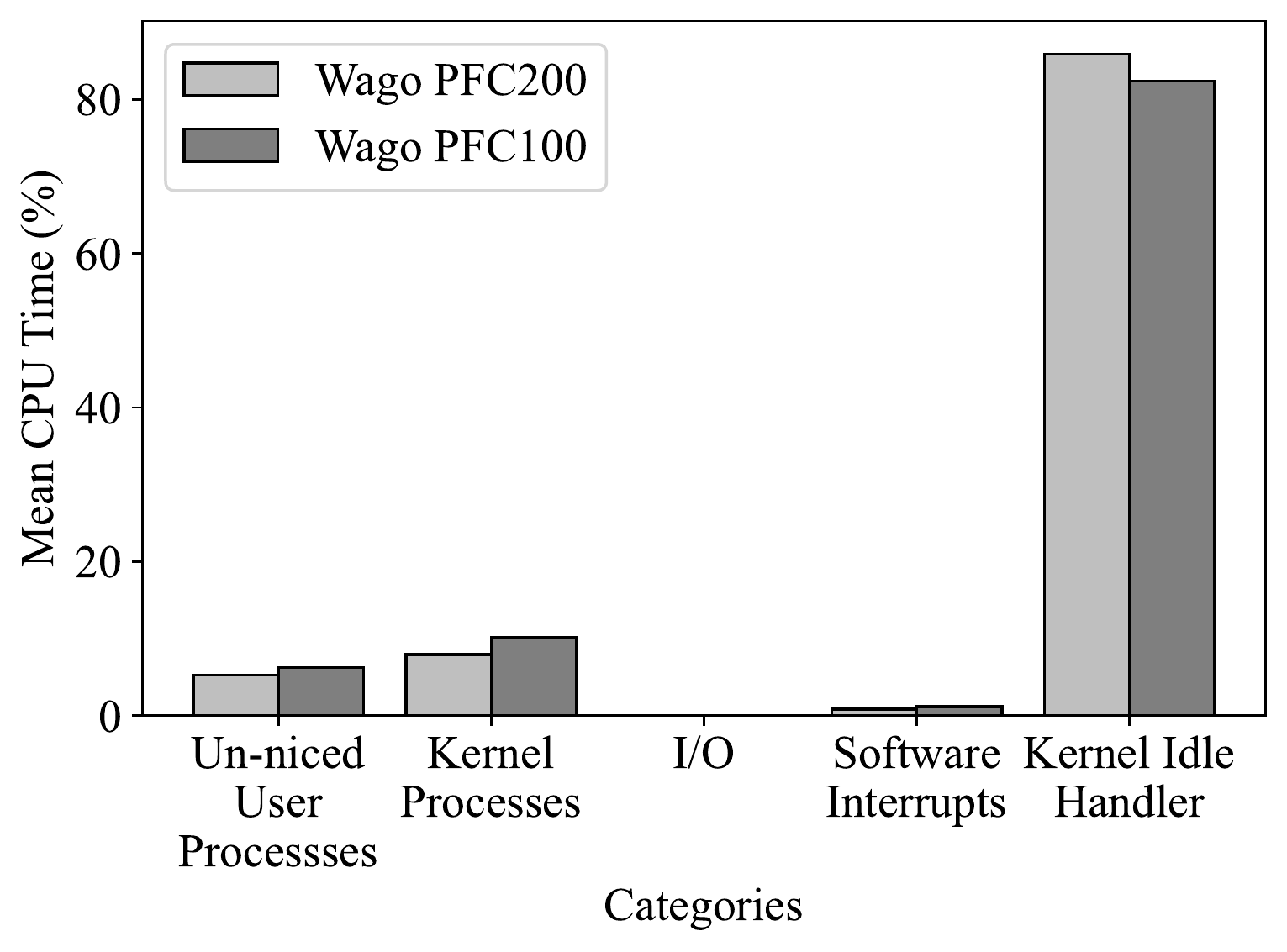}
  \caption{An overview of CPU utilization by different operations on a PLC.}\label{fig:latency_overview}
\endminipage\hfill
\minipage{0.485\columnwidth}%
  \includegraphics[width=\linewidth, trim={0.2cm 0.3cm 0.1cm 0.2cm}, clip]{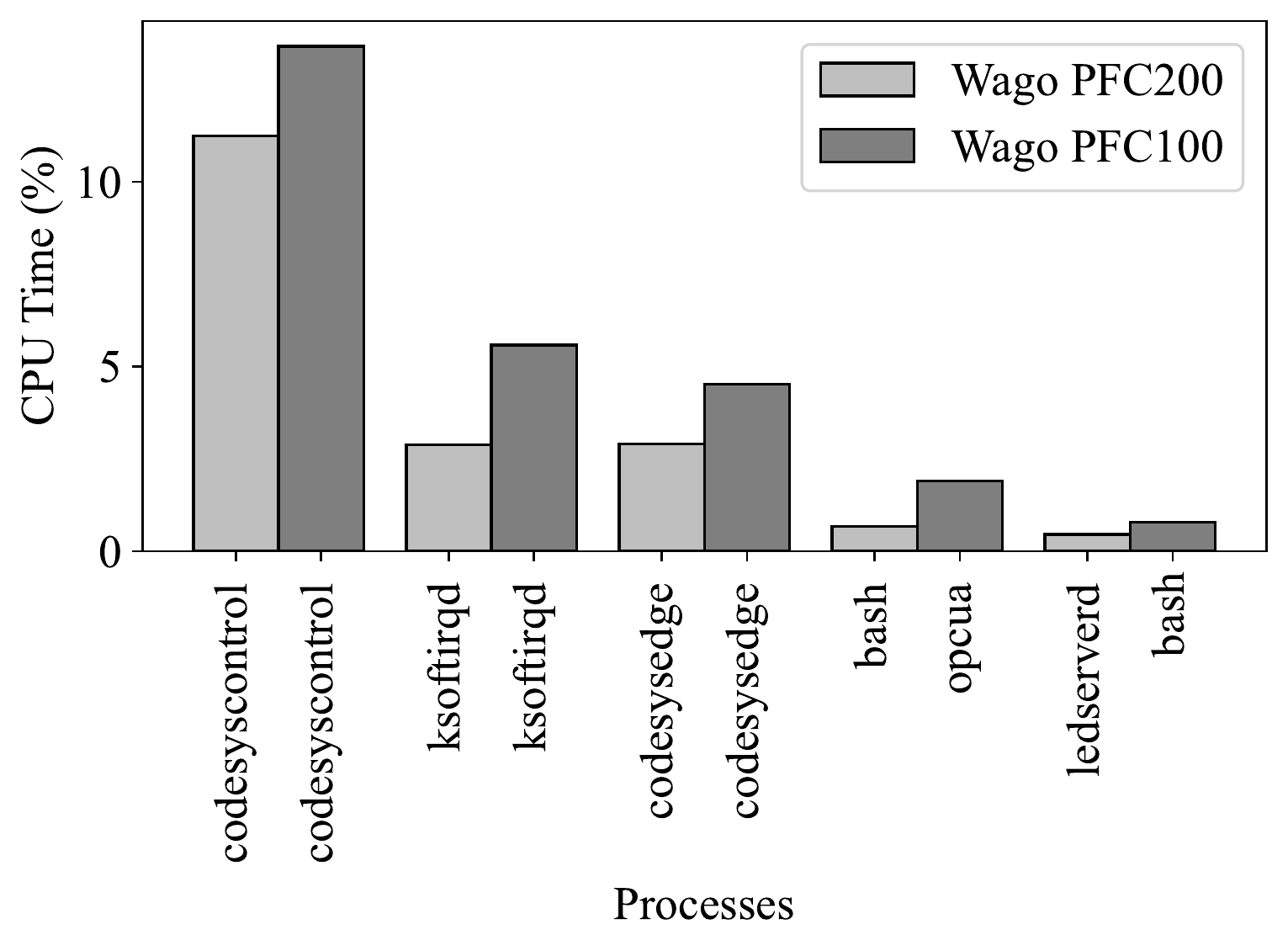}
  \caption{CPU utilization of the top five processes on a PLC.}\label{fig:top5_latency}
\endminipage
\vspace{-0.35cm}
\end{figure}

\noindent{\bf Non-intrusiveness.} ICSPatch utilizes an LKM-based patcher for writing the patch at appropriate memory locations. Inserting the hook (modified \texttt{ldr} instruction) is the most critical part of the patching process, resulting in a modified execution flow toward the patch. Since PLCs often execute straightforward dedicated process logic, they lack heavy computation power. For instance, Wago PFC100 and PFC200 (Generation 2) utilize single-core Cortex-A8 processors. To execute multiple processes, such processors utilize time slicing. As discussed earlier, the LKM patcher inserts the patch when the runtime process is switched out and sleeping. To further ensure that patching commences safely in an atomic fashion, the runtime process priority is adjusted, and our LKM temporarily disables kernel preemption and interrupt handling. 

We write a \texttt{bash} script for Wago PFC100 and PFC200 to measure the execution and sleeping time for the runtime. We measure running/sleeping times in an interval of 1 hour (3600 seconds). We observe that, on average, for a basic out-of-the-box installation, the runtime process utilizes $\approx$492 s ($\approx$13.66\%) and $\approx$404 s ($\approx$11.23\%) of total CPU time for Wago PFC100 and PFC200, respectively, while executing the control application binary for regulating the steam flow into the brine heater for thermal desalination plants, explained further in Section~\ref{sec:case_study}. While the \texttt{codesyscontrol} process still occupies the majority of the CPU time when compared to any other processes, as shown in Figure~\ref{fig:top5_latency}, the majority of the time is spent in the kernel Idle handler, $\approx$2966 s ($\approx$82.41\%) and $\approx$3092 s ($\approx$85.6\%) for Wago PFC100 and PFC200 respectively, as shown in Figure~\ref{fig:latency_overview}. This affirms that the LKM patcher can utilize this idle time on these single-core PLC devices to non-intrusively patch the currently executing control application binaries.

\section{Case Study: Hotpatching Out-of-bounds Write Vulnerability in a Desalination Plant} \label{sec:case_study}
\begin{figure}[t]
    \centering
    \includegraphics[width=\columnwidth, trim={5.6cm 8.1cm 1.6cm 3.1cm}, clip]{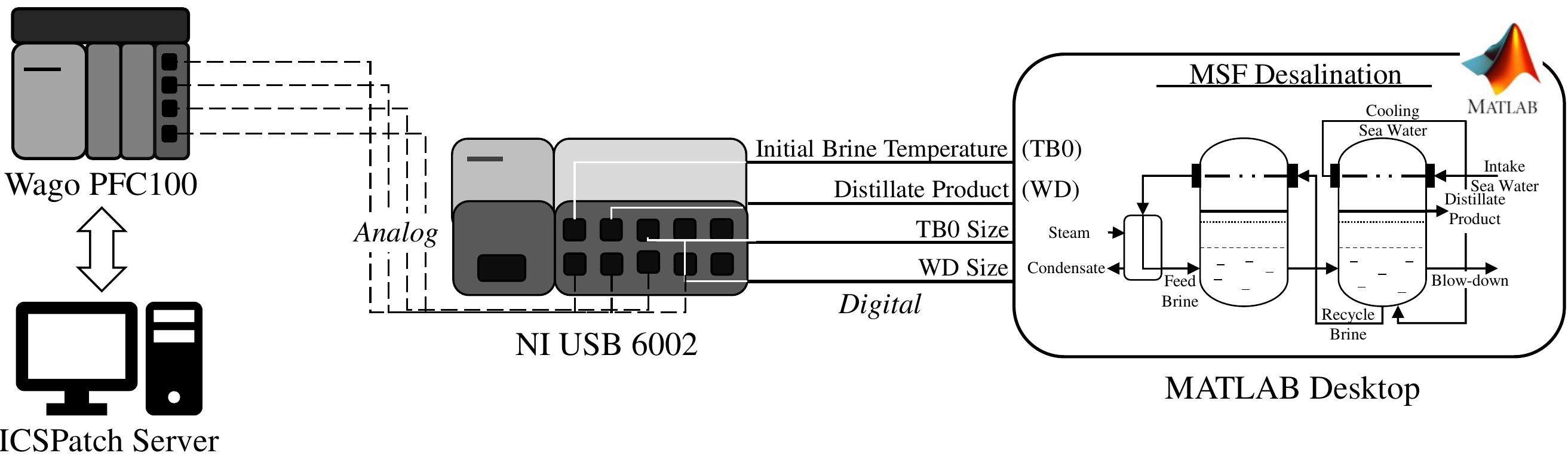}
    \captionof{figure}{Experimental setup for patching out-of-bound write vulnerability in an MSF desalination plant HIL.}
    \label{fig:case_study}
    \vspace{-0.35cm}
\end{figure}

\noindent \textbf{Hardware-in-the-loop setup.} For testing the applicability of ICSPatch on PLCs deployed in the field, monitoring and regulating physical processes in critical infrastructure, we perform a Hardware-in-the-Loop (HIL) experiment. The experiment setup is shown in Figure~\ref{fig:case_study}, consisting of a MATLAB Simulink model for a Multi-Stage Flash desalination plant validated against the Khubar II plant in Saudi Arabia and utilized for HIL experiments in literature~\cite{proc:msf_process_aware, art:icsml, proc:jtag_malware_1, proc:jtag_malware_2}. The model controls the steam flow into the heating section based on the initial brine temperature (TB0) and the distillate output (WD). It runs on a host connected to NI USB 6002, a data acquisition (DAQ) device providing multiple digital and analog I/Os, which converts the digital inputs from the Simulink model to analog outputs and then feeds them to the Wago PFC100 PLC. The PLC sends the output of the control logic as analog values back to the DAQ device, which converts it to digital values that the Simulink model reads. We extract TB0 and WD from the Simulink model and send them to the PLC. ICSPatch server is connected via SSH to the PLC.

\begin{figure}[t]
    \centering
    \vspace{-0.1in}
    \includegraphics[width=\columnwidth, trim={0.2cm 0.2cm 0.2cm 0.2cm}, clip]{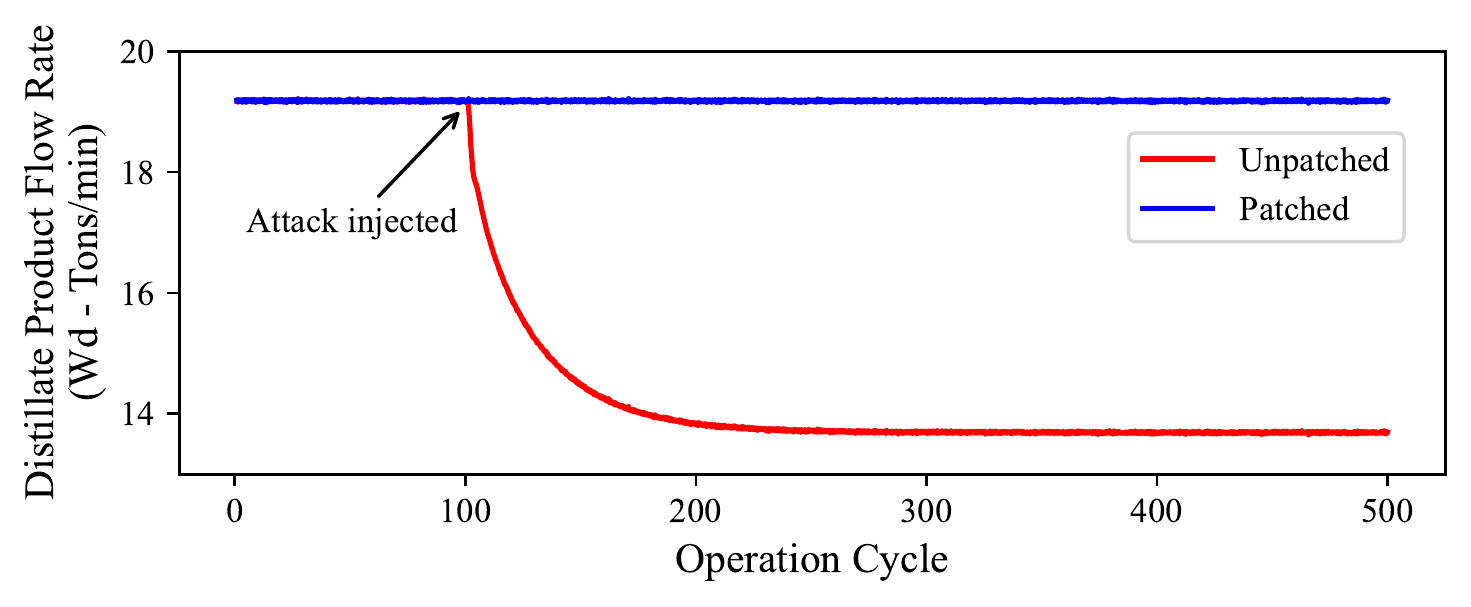}
    \captionof{figure}{Distillate product flow rate for control application with unpatched and patched out-of-bounds write vulnerability.}
    \label{fig:HiL_Result}
    \vspace{-0.35cm}
\end{figure}

\noindent \textbf{Out-of-bounds write.} The control logic executing on Wago PFC100 and regulating the critical physical process utilizes \texttt{MemUtils.MemCpy}, an imported function for copying memory contents into temporary variables in a function. An adversary can maliciously modify the bound value to trigger an out-of-bounds write on critical runtime variables on the stack, resulting in a crash. In our example, as shown in Figure \ref{fig:HiL_Result}, an adversary triggers an out-of-bounds write attack at the 100$^{th}$ cycle of the model by maliciously modifying the TB0 size and overwriting critical sections of the runtime stack.

\noindent \textbf{Patching.} ICSPatch extracts memory segments from the development PLC loaded with the exploit input, rehosts them in \texttt{angr}, and performs concolic execution to detect vulnerabilities. Once it detects the vulnerability, it localizes it, selects a patch location, connects with the deployment PLC, and patches it on the deployment PLC by adding a bound check to the size of the TB0 parameter, which is the root cause of the vulnerability in the control logic.

Figure~\ref{fig:HiL_Result} shows the consequences of unpatched control logic regulating the MSF process (shown in red) with an out-of-bounds write vulnerability. The output Wd flow rate decreases significantly due to the attack beginning on the 100$^{th}$ operational cycle of the simulation. The adversary crashes the runtime by triggering an out-of-bounds write vulnerability. However, a patched instance of the control application is protected against the attack, and the control logic functions as expected without any output changes.

\section{Scalability Discussion} \label{sec:discussion}
\textbf{Localizing and patching more vulnerabilities.} We include eight different memory-related functions in our dataset and successfully patch the vulnerable control binaries using ICSPatch. The patches for these memory-related functions implement a bound check and are therefore applicable to all the relevant examples with minor modifications to the live memory addresses of the deployed PLC. In most cases, extending ICSPatch to support other functions requires adding the appropriate node to the DDG for automated vulnerability localization.

\noindent \textbf{Hotpatch deployment.} ICSPatch requires a kernel-level component (LKM, in our case, or a dedicated patch driver) for patching the control applications running on Linux operating systems. However, PLCs may run bare-metal firmware. In that case, ICSPatch also provides a JTAG-based patcher out-of-the-box to support such devices, which we have successfully tested on a BeagleBone Black (BBB). Similarly, HERA implements a low-priority task, and RapidPatch assumes a traditional OTA transfer tunnel for its patch deployment.

\noindent \textbf{Targeting more devices.} We extend ICSPatch to work with Codesys runtime running on BBB. The main differences appear in the methodology used for branching between functions. The Codesys variant for Wago PFC100 saves the \texttt{R14} and explicitly modifies \texttt{PC} (Figure~\ref{fig:codesys_wago_exec_flow}), whereas, on BBB a \texttt{blx} (branch with link exchange) instruction is used for branching to the next function. Due to the discrepancies in branching, the patch requires minor modifications to accommodate the use of different registers and branching mechanisms. Furthermore, the runtime variant operating across devices might also vary in the mode of its compilation. For instance, the runtime for BBB is compiled in Thumb mode, whereas it is ARM for Wago PFC100. Therefore a branch to the following function on BBB expects the code to be in Thumb, whereas it has to be in ARM on PFC100, requiring the patches to be compiled accordingly.

\begin{table}
\centering
\caption{Firmware base and runtime support of popular ICS vendors.}
\label{table:ics_codesys_vendor}
\setlength\tabcolsep{13pt}
\resizebox{\columnwidth}{!}{
\begin{tabular}{ccc} 
\toprule
\multicolumn{1}{c}{\textbf{Vendor}} & \textbf{Firmware Base} & \textbf{Supports Codesys}             \\ 
\midrule
ABB                                 & Linux                  & Yes  \\
Beckhoff                            & FreeBSD                & Yes                      \\
Bosch Rexroth                       & Linux                  & Yes                       \\
Mitsubishi                          & N/A                    & Yes                      \\
Rockwell Automation                 & VxWorks                & No                     \\
Schneider Electric                  & VxWorks                & Yes                      \\
Siemens                             & OpenBSD                & No                     \\
WAGO                                & Linux                  & Yes  \\
\bottomrule
\end{tabular}}
\vspace{-0.35cm}
\end{table}

ICSPatch can be extended to support other vendors, including Siemens, Rockwell, Mitsubishi, and Schneider, who collectively share approximately 75\% of the PLC market \cite{misc:statistaPLCmarket}. While older PLCs primarily used monolithic proprietary firmware containing only in-house developed components, this is no longer the case, as shown in Table~\ref{table:ics_codesys_vendor}. Instead, the ICS market has transitioned to a model that utilizes commodity operating systems and third-party components, enabling the deployment of tools like ICSPatch. For instance, our analysis of other PLC firmware reveals that the Siemens SIMATIC S7 line utilizes OpenBSD while Rockwell Automation and Schneider Electric use the VxWorks kernel. Looking closer, we noticed that the firmware bundles had numerous proprietary and third-party open-source components. Such components included JavaScript libraries like jQuery, utilities like OpenSSL, ICS-specific software like the Codesys runtime and Unified Automation's OPC UA SDK, and even more bespoke dependencies like the Dinkumware C++ framework, IBM Rational Rhapsody, and the Electronic Arts standard template library. Beyond the underlying OS and platform support, ICSPatch also depends on the disassembly of the control binary during the parsing phase. Manufacturers like Mitsubishi, Beckhoff, and Schneider Electric can be easily supported since their PLC ecosystems integrate the Codesys runtime. PLCs from other vendors like Siemens can be supported using the corresponding tools in their respective ecosystems, like the JEB S7 PLC Block Decompiler \cite{misc:jebS7} or the Rizin framework \cite{misc:rizinS7}.

\section{Limitations}
\label{sec:limitations}
\noindent \textbf{Codesys specific implementation.} As mentioned, ICSPatch has three main components: Vulnerability identification, localization, and hotpatching. Vulnerability localization based on DDG does not assume Codesys-specific compilation and can work with other platforms. Likewise, modifications to the \texttt{angr} rules can also identify vulnerabilities in control applications compiled for other platforms. However, the patches are designed with the Codesys compiler as a target. Therefore, some effort is required to create appropriate skeleton patches to extend applicability to other platforms.

\noindent \textbf{Patching multiple vulnerabilities.} A control application may have multiple vulnerabilities; for instance, the same malicious input (bound) utilized by two or more \texttt{SysMem.SysMemSet} functions for initializing memory. However, in its current version, ICSPatch only patches the first instance of the vulnerability.

\noindent \textbf{Requiring user input for patching.} While, for the most part, ICSPatch is automated to identify, locate, and deploy the patch, it cannot automatically suggest an appropriate bound check for memory-related vulnerabilities (for out-of-bounds read/write), thus requiring user input during patch generation. Nevertheless, patching OS command injection vulnerabilities is entirely automated as it involves fixing the address table.

\noindent \textbf{Patching on multicore devices.} ICS devices, for the most part, do not employ powerful multicore processors as they execute simple and definite logic. The non-intrusiveness of ICSPatch relies on the single core execution of all processes. A more elaborate mechanism for ensuring non-intrusiveness would be required in multicore cases.

\section{Related Work} \label{sec:related_work}
\noindent \textbf{Automated vulnerability identification and localization.} Machine Learning can be used for vulnerability localization; for instance, Rebecca et al. utilize deep feature representation learning on lexed source code~\cite{proc:auto_vuln_deep_representation}. Devign employs graph-level classification with a graph neural network-based model to extract valuable features learned from rich node representations~\cite{art:devign}. On the other hand, for binaries, Runhao et al. propose a gradient-guided vulnerability localization method: DeepVL, utilizing execution traces to filter vulnerable basic blocks~\cite{art:dnn_binaries_localization}. Finally, Shiqi et al. utilize statistical localization technique to discover the root cause of the bug, given just one exploit input~\cite{proc:vuln_localization}.

\noindent \textbf{Hotpatching in Android.} Yue et al. put forth KARMA, a system that produces patches written in high-level memory-safe language applicable at multiple levels in the kernel to enable malicious input filtering~\cite{proc:adaptive_live_patching}. VULMET utilizes weakest precondition reasoning to modify official patches into semantically correct hotpatches~\cite{proc:anroid_patching_2}. Yue et al. propose LIBBANDAID for automatically generating updates for third-party libraries~\cite{proc:third_party_patches}. Finally, bowknots for kernel bugs nullify the side effects of currently executing syscalls triggering bugs~\cite{proc:kernel_patching_1}.

\noindent \textbf{Hardware-based hotpatching.} Traditional hotpatching employs trampolines; however, approaches such as InstaGuard use basic debugging primitives supported by ARM CPUs to enable a rule-driven hotpatching mechanism without injecting any code~\cite{proc:instaguard}. HERA, supported on ARM Cortex-M3/M4 devices, utilizes its Flash Patch and Breakpoint (FPB) unit to redirect the vulnerable code to the patch~\cite{proc:hera}. On the other hand, a recently proposed technique, RapidPatch, facilitates a dynamic code replacement technique utilizing eBPF virtual machines for executing patches on an embedded device with resource constraints~\cite{proc:rapid_patch}.

\noindent \textbf{Comparison to state-of-the-art.} HERA and RapidPatch are the current state-of-the-art hotpatching solutions for RTOSes employed for embedded devices, thus, the closest to our work. HERA utilizes FPB in Cortex-M3/M4, and RapidPatch employs eBPF virtual machines. However, these solutions cannot directly be employed for PLC control binaries because:
\begin{itemize}[nosep, leftmargin=0.05in]
\item Control applications compiled into nonstandard file format run in the context of a proprietary piece of software, the runtime. So, hotpatching requires understanding control application internals while handling runtime abstractions.
\item Vulnerabilities manifest in the imported functions, and due to their shared nature, they require patching within the boundaries of the control application, necessitating localization.
\item Unlike the assumption made by the current state-of-the-art techniques, ICSPatch does not have access to any upstream patch source for creating the hotpatch.
\item HERA utilizes hardware-specific debugging features for redirecting execution flow to the patch (Cortex-M3/M4), limiting its applicability. For instance, Wago PFC100 and PFC200 employ Cortex A8, lacking the FPB debug unit.
\item RapidPatch utilizes ePBF virtual machines, which cannot directly be applied for control applications as they execute inside the runtime with the proprietary file format.
\end{itemize}

\section{Conclusion} \label{sec:conclusion}
This work proposes ICSPatch, a tool for automated vulnerability identification by detecting violations to security specifications, localization by traversing a DDG, and hotpatching using an LKM-based patcher. We implement ICSPatch for the Codesys platform, deployed on over 400 known ICS devices from 80 industrial device vendors. We successfully patch out-of-bounds write/read, OS command injection, and invalid input validation vulnerabilities spread across a dataset of 24 synthetic control application binaries while only incurring negligible execution and memory overheads. We also demonstrate ICSPatch patching a live PLC controlling a HIL simulation of an industrial process.

\section*{Acknowledgment}
This research was supported by the NYU Abu Dhabi Global Ph.D. Fellowship. ICSPatch and the dataset can be found at \href{https://github.com/momalab/ICSPatch}{{github.com/momalab/ICSPatch}}.

\clearpage


\bibliographystyle{plain}
\bibliography{references}

\end{document}